\documentclass[
aps,
prd, 
twocolumn,
superscriptaddress,
nofootinbib,
amsmath,amssymb,amsfonts,
]{revtex4-2}

\usepackage{aas_macros}
\usepackage[T1]{fontenc}
\usepackage[english]{babel}
\usepackage{bm} 
\usepackage{empheq}
\usepackage{xcolor}
\usepackage[breaklinks=true,colorlinks,citecolor=blue,linkcolor=blue,urlcolor=blue]{hyperref}

\usepackage{subcaption} 

\usepackage{enumitem} 

\usepackage{amsmath,amsthm,amssymb} 

\usepackage{mathtools,bbm}

\usepackage{starfont} 
\DeclareSymbolFont{starfontsym}{OT1}{sts}{m}{n}
\DeclareMathSymbol{\ascnode}{\mathord}{starfontsym}{107}

\newcommand{\hampaper}{Ref.~1}

\newcommand{\naturals}{\mathbb{N}} 
\newcommand{\reals}{\mathbb{R}} 
\newcommand{\complexes}{\mathbb{C}} 
\newcommand{\ee}{\mathrm{e}} 
\newcommand{\ii}{\mathrm{i}} 

\let\Re\relax
\DeclareMathOperator{\Re}{\mathfrak{Re}} 
\let\Im\relax
\DeclareMathOperator{\Im}{\mathfrak{Im}} 

\renewcommand{\vec}[1]{\boldsymbol{#1}} 
\newcommand{\uvec}[1]{\hat{\vec{#1}}} 
\newcommand{\mat}[1]{\mathbf{#1}} 

\newcommand{\idmat}[1]{\mathbbm{1}_{#1}} 
\DeclareMathOperator{\diag}{diag} 
\DeclareMathOperator{\antidiag}{antidiag} 

\newcommand{\tr}[1]{\prescript{t}{}{#1}} 

\newcommand{\lagstd}{\mathcal{L}} 
\newcommand{\hamstd}{\mathcal{H}} 
\newcommand{\lagext}{\Lambda} 
\newcommand{\hamext}{\mathcal{A}} 

\newcommand{\labelA}{{\uparrow}} 
\newcommand{\labelB}{{\downarrow}} 
\newcommand{\pathA}[2][]{{#2}_{\labelA #1}} 
\newcommand{\pathB}[2][]{{#2}_{\labelB #1}} 
\newcommand{\pathAB}[1]{{#1}_{\uparrow\joinrel\downarrow}} 

\newcommand{\pathmetric}[1][ab]{\eta_{#1}} 

\newcommand{\PL}{\mathrm{PL\!}}
\newcommand{\vorder}[1]{\mathcal{O}(-^{#1})} 
\newcommand{\porder}[1]{\mathcal{O}(\varepsilon^{#1})} 

\DeclareMathOperator{\dilog}{Li_2} 

\newcommand{\avg}[1]{\left\langle {#1} \right\rangle} 
\newcommand{\primitive}{\mathfrak P} 
\newcommand*{\argdot}{\makebox[1ex]{\textbf{$\cdot$}}} 
\DeclarePairedDelimiter\norm{\lVert}{\rVert}%

\newcommand{\dd}{\mathrm{d}} 
\newcommand{\grad}{\vec{\nabla}} 
\newcommand{\dext}[2]{\frac{\partial #1}{\partial #2\!\!}\,} 

\newcommand{\lPB}{\{\!\{}
\newcommand{\rPB}{\}\!\}}

\let\originalleft\left
\let\originalright\right
\renewcommand{\left}{\mathopen{}\mathclose\bgroup\originalleft}
\renewcommand{\right}{\aftergroup\egroup\originalright}
\DeclarePairedDelimiter\pars{\lparen}{\rparen}

\begin{document}

\preprint{APS/123-QED}

\title{Nonconservative Lie series: \\ post-Newtonian binary dynamics at 2.5PN}

\author{C.~Aykroyd}\email{christopher.aykroyd@obspm.fr}
\affiliation{LTE, Observatoire de Paris, Universit\'e PSL, Sorbonne Universit\'e, Université de Lille, LNE, CNRS, 61 avenue de l'Observatoire, 75014 Paris, France}

\author{A.~Bourgoin}
\affiliation{LTE, Observatoire de Paris, Universit\'e PSL, Sorbonne Universit\'e, Université de Lille, LNE, CNRS, 61 avenue de l'Observatoire, 75014 Paris, France}

\author{C. Le Poncin-Lafitte}
\affiliation{LTE, Observatoire de Paris, Universit\'e PSL, Sorbonne Universit\'e, Université de Lille, LNE, CNRS, 61 avenue de l'Observatoire, 75014 Paris, France}


\date{\today}

\begin{abstract}
    We develop a Lie series perturbation method for solving the dynamics of nonconservative systems within the doubled phase-space Hamiltonian framework introduced in our previous work [Phys.~Rev.~E 113, 034137]. The procedure can be applied to a wide range of perturbed nonconservative systems which are not initially given by a Hamiltonian, by casting them into the doubled phase-space where the Lie method can be employed. 
    As an application, we tackle the non-spinning $2.5$ post-Newtonian binary problem at leading dissipative order. Starting from the ADM radiation-reaction dynamics, we construct a resonant Birkhoff normal-form and the corresponding generator of the canonical transformation. Next, we establish an analytical perturbative solution accounting for both the long-term (secular) and short-term (oscillatory) temporal behavior, without performing a small-eccentricity expansion, for nondegenerate bound eccentric orbits. The secular piece is encoded in a resonant normal-form, and captures the long-term inspiral, reconstructing exactly the known Peters-Mathews relations for semi-major axis and eccentricity. The oscillatory piece completes the dynamics by supplying instantaneous, non-averaged corrections to the orbital parameters.
    Combined, the two pieces give a controlled, fully canonical description of the dissipative dynamics. 
\end{abstract}


\maketitle


\section{Introduction}

\subsection{Context}

Standard Lagrangian and Hamiltonian mechanics are not adequately equipped for dealing with nonconservative processes. While dissipation can be easily incorporated at the level of the equations of motion, deriving these same dynamics from a variational principle is far more challenging. From a Lagrangian standpoint, several workarounds are known. First, time-dependent Lagrangians are system-specific and do not provide sufficient flexibility for encoding arbitrary dissipative dynamics \cite{Riewe1996}. Other options include the use of convolution products in the action \cite{Gurtin1963, Gurtin1964, Tonti1973} (which limits the dynamics to linear equations) or fractional derivatives \cite{Riewe1996, Riewe1997, Dargush2012, Dargush2012a} (at the cost of operator non-locality). On the Hamiltonian side, contact mechanics \cite{Bravetti2017} encodes dissipation geometrically on a contact manifold. A common takeaway \cite{Gurtin1963, Tonti1973, Dargush2012, Galley2013, Riewe1996} is that classical action principles, which are formulated as boundary problems, naturally suppress irreversibility, which is a characteristic of dissipative effects; in conclusion, an appropriate description of dissipative evolution requires an action compatible with initial value problems (IVP).

One way to produce such IVP-compatible actions is through the introduction of extra degrees of freedom. A longstanding example was provided by Bateman \cite{Bateman1931}, who tackled linear dissipative problems. Bateman coupled each dissipative degree of freedom to an auxiliary variable; the auxiliary variables evolved according to an adjoint system, which absorbed energy dissipated by the main system. Thus, the combined system was conservative and fit within the traditional framework. Although Bateman's approach was originally limited to linear systems, it paved the way for the development of more sophisticated strategies. Indeed, in the context of quantum field theory, Schwinger \cite{Schwinger1961} and Keldysh \cite{Keldysh1964} constructed a powerful technique for studying out-of-equilibrium systems via path-integral methods. More recently, Galley and collaborators formulated the classical (i.e., non-quantum) analog \cite{Galley2013, Galley2014,Aykroyd2025}, an action principle based on initial data. In this formalism, the dimension of the configuration space is duplicated: each degree of freedom gets replaced by two copies, $\vec q \to (\pathA{\vec q}, \pathB{\vec q})$, representing the same underlying physical quantity evolving along identical trajectories. One then builds a Lagrangian 
\begin{align}
    \lagext(\pathA{\vec q}, \pathB{\vec q}, \pathA{\dot{\vec q}}, \pathB{\dot{\vec q}}) & = \lagstd(\pathA{\vec q}, \pathA{\dot{\vec q}}) -  \lagstd(\pathB{\vec q}, \pathB{\dot{\vec q}}) \nonumber\\
    & +\mathcal K(\pathA{\vec q}, \pathB{\vec q}, \pathA{\dot{\vec q}}, \pathB{\dot{\vec q}}),
\end{align}
which is antisymmetric with respect to a swap of the two paths. Here, $\lagstd$ produces the conservative dynamics and $\mathcal K$ couples the two paths, thereby encoding nonconservative effects. With suitable initial data, the Euler-Lagrange equations of $\lagext$ admit a unique solution in which the two paths $\pathA{\vec q}$ and $\pathB{\vec q}$ collapse into a single physical trajectory $\vec q$ (see \cite{Aykroyd2025} and Fig. 1 therein); under this condition, the equations reduce to the standard conservative dynamics with additional nonconservative forces built from derivatives of $\mathcal K$. 
In this setup, the doubled variables should not be interpreted as two independent physical binaries, but as a mathematical construct introduced so that an initial-value problem can be obtained from a stationary-action principle. Off-shell, the two paths are allowed to vary independently; on-shell, uniqueness of the initial-value problem and the label-exchange antisymmetry force them to coincide.

In our previous work ({\hampaper} \cite{Aykroyd2025}), we leverage Galley's IVP action principle to construct the corresponding Hamiltonian formulation for nonconservative systems. Within this formulation, the dynamics of a system is embedded into a symplectic manifold of doubled dimension, evolving under the flow of a Hamiltonian which is antisymmetric with respect to the exchange of the two paths. Essentially any classical discrete IVP can be described within the framework; we construct explicit extended Hamiltonians encoding arbitrary second-order ordinary differential equations.
In the absence of dissipation, our formulation reduces to two isolated copies of phase-space possessing identical dynamics, providing a direct bridge to standard Hamiltonian mechanics. In the presence of dissipation, the two copies of phase-space become coupled, and the true physical dynamics arises on a smaller dimensional slice of the doubled manifold.
In practice, one replaces a dissipative force law, which is not Hamiltonian on the ordinary physical phase space, by an autonomous Hamiltonian flow on an enlarged symplectic space. The usual canonical tools, such as Poisson brackets, canonical transformations, Hamilton-Jacobi, and Lie series, can then all be applied before taking the physical limit. For PN applications this is useful because the conservative sector is already naturally expressed in the Hamiltonian formalisms of Arnowitt-Deser-Misner (ADM) and the effective one-body (EOB), while the radiation-reaction sector is usually appended at the level of the equations of motion or through explicitly time-dependent Hamiltonians. Instead, the doubling of variables allows a description of the conservative and dissipative dynamics within a single unified structure.


In the present work, we exploit our Hamiltonian framework to develop an approach for solving the dynamics of nonconservative systems within a canonical setting.
Gravitationally radiating binaries are an especially compelling subject of application; the radiation reaction is a fundamentally nonconservative force which cannot be easily accommodated within standard Hamiltonian treatment. Gravitational waves radiate away energy and angular momentum from the binary, causing orbits to inspiral and circularize. In post-Newtonian (PN) frameworks \cite{Blanchet2024, Schafer2024}, the radiation reaction forces first arise at the 2.5PN order, and have been obtained in the literature through three different methods.
The first technique performs a direct integration of the retarded field generated by the source to obtain the complete equations of motion; in this way the leading dissipative terms were obtained at 2.5PN and 3.5PN order \cite{Damour1981,Kopejkin1985,Blanchet1998,PatiWill2002}. The second method relies on a matching between the fields in the near-zone and the wave-zone, or equivalently on radiation-reaction potentials constructed from the matched exterior field; this has been achieved up to 4PN \cite{Blanchet1984, Blanchet1997, Nissanke2005}, where hereditary tail effects enter, reflecting the back-scattering of outgoing waves off the background curvature. The third approach invokes a balance between the energy and angular momentum fluxes radiated by gravitational waves at infinity and the local losses in the binary equations of motion; this was extended up to 4.5PN order \cite{Iyer1993,Iyer1995,Gopakumar1997,Gopakumar1997b}. The resulting dissipative forces are generally inserted directly into the equations of motion. Alternatively, in the ADM and EOB Hamiltonian formalisms, the radiation reaction has been described through an explicitly time-dependent Hamiltonian: in ADM at 2.5PN and 3.5PN order \cite{Schafer1985,Jaranowski1997,Konigsdorffer2003}, and in the effective-one-body framework through the incorporation of the corresponding radiation-reaction information into the effective dynamics \cite{Buonanno1999}.

The next step is to extract \emph{solutions} to the radiative dynamics, which has been achieved through a variety of formalisms. Early influential treatment was brought by \citet{Peters1963}, who computed orbit-averaged evolution equations for the semi-major axis and eccentricity. Since then, the full (beyond orbit-averaged) solutions have been computed. \citet{Damour2004} employed the method of variation of arbitrary constants, treating the binary acceleration as a leading-order correction to the conservative motion described through the quasi-Keplerian coordinates \cite{Damour1985, Memmesheimer2004}. The method was then extended to $3.5$PN by \citet{Konigsdorffer2006}. Other work by \citet{Zixin2019} alternatively solved the $2.5$PN radiation reaction plus leading-order spin-orbit effects through the dynamical renormalization group approach.

These solutions are useful not only for studying the binary inspiral, but they can also potentially be used in waveform applications to produce analytical templates from a canonical route.
Indeed, the post-Newtonian wave generation formalism relates the gravitational waves reaching a detector in the far-zone to the stress-energy of the source, and can be split into multiple steps. First, one solves the near-zone dynamics of the compact binary, obtaining analytical evolutions for the relative position, velocity, and when relevant, spin variables. Next, these dynamical variables are substituted into the source/radiative multipole moments, from which the far-zone gravitational waveform is constructed, as outlined below.

The metric perturbation $h_{ij}^{\rm TT}$ in the far zone can be decomposed into a set of radiative mass and current multipole moments $(\mathcal U_L, \mathcal V_L)$ \cite{Blanchet1989}. At Newtonian order the waveform depends only on the second time derivative of the mass quadrupole, but a $2.5$PN accurate orbital dynamics requires that the waveform be computed to higher PN order consistently. A table of the required orders for each multipole moment is summarized for instance in \citet{Kidder2008}. Beyond the instantaneous terms, hereditary contributions such as tails appear from relative $1.5$PN order onward \cite{Blanchet1992}, with nonlinear memory effects as hereditary integrals in the radiative-moment construction formally at $2.5$PN \cite{Favata2010}.
These radiative multipole moments can be related to six sets of \emph{source}/gauge multipole moments $\big( \mathcal I_L, \mathcal J_L, \mathcal W_L, \mathcal X_L, \mathcal Y_L, \mathcal Z_L \big)$ by applying the far-zone/near-zone matching of the multipolar post-Minkowskian formalism \cite{Blanchet2006, Blanchet2024}. Typically, a set of two canonical source moments $(\mathcal M_L, \mathcal S_L)$ is computed as an intermediate step. The source multipole moments themselves originate from the stress-energy pseudo-tensor, including both matter and gravitational field contributions. For compact binaries modeled as point-masses, these moments can be evaluated as function of the binary masses and the dynamical variables in the center-of-mass frame. 

Once an analytical solution for the radiative orbital dynamics has been obtained, it can be inserted into these multipolar expressions to construct the corresponding waveform \cite[see e.g.][]{Arun2004}. The evaluation of hereditary tail and memory integrals requires special treatment, see \cite{Arun2004} for the case of circular orbits. Finally, the resulting $h_{ij}^{\rm TT}$ is contracted with the polarization vectors to obtain the plus and cross polarizations $h_+$ and $h_\times$. In this sense, improvements to the solutions for the radiative dynamics directly reflect in the waveform. 

It is important to note that since the wave-generation formalism is typically carried out in harmonic coordinates, one might need to convert between harmonic and ADM if the solutions are computed in a canonical formalism. Explicit coordinate transformation relations are available through 3PN \cite{Damour2001}, not accounting for the dissipative part of the dynamics. At 2PN, the waveform has already been expressed in ADM coordinates in \cite{Gopakumar2002}.

\subsection{Plan of the paper}

The aim of this paper is to show that the nonconservative Hamiltonian framework can be combined with perturbation methods to produce analytical solutions for dissipative systems. We are particularly interested in extending the Lie series method, a fully canonical perturbation technique with several advantageous features. In essence, the Lie series provides a formal way of averaging some dynamics over one period of the unperturbed motion. It gives an algorithmic construction which separates the solution into a secular (orbit-averaged) component and oscillatory (non-averaged) corrections which are often neglected in other averaging methods. Lie methods also avoid unbounded Poisson-type\footnote{By ``Poisson-type terms'' we mean terms of Poisson series form, namely trigonometric contributions multiplied by polynomial factors in time or in angles (for instance $t \sin{t}$). In standard perturbation techniques, such terms can emerge as solutions to purely oscillatory dynamics but behave secularly over long time scales.} terms that can arise in standard perturbation techniques, and more directly resemble a Fourier-like expansion with frequencies corrected order-by-order in the expansion parameter. Furthermore, it allows for a systematic search of resonances and hence allows one to discuss the integrability of a given system. 

We therefore generalize the Lie series perturbation technique to nonconservative systems.
This new procedure can be generally applied to solve the dynamics of any perturbed integrable system once they have been cast into double-variable Hamiltonian form. In practice, this includes systems which are not initially given by a Hamiltonian, which can be put into Hamiltonian form through the canonicalization procedure discussed in {\hampaper} \cite{Aykroyd2025}.

As a test case, we apply the nonconservative Lie series to solve the $2.5$PN radiation reaction dynamics. The foundation is the conservative Lie series treatment for rotation-invariant Hamiltonians set up in our other previous work \cite{Aykroyd2024}, where we built explicit generators and the complete time-domain solutions for post-Newtonian spinless compact binaries at $2$PN. Here, we extend this strategy to a dissipative setting.

In the context of Hamiltonian ADM/EOB methods, we also provide a recipe for encoding the dissipative sector through a fully canonical description. 
Standard ADM treatments of radiation reaction often introduce a time-dependent Hamiltonian whose expression is dependent on the orbital solution. Structurally, the Hamiltonian can be expressed as:
\begin{equation}
    \mathcal H(\vec q, \vec p, t) = f(\vec q, \vec p, \vec q(t), \vec p(t)) .
\end{equation} 
The second occurrence of $\vec q$ and $\vec p$ in the arguments of $f$ needs to be replaced by explicit time solutions to the dynamics, $\vec q(t)$ and $\vec p(t)$, so that $f$ is implicitly a function of time. Thus, the time-dependence of the Hamiltonian is not explicitly known before solving the dynamics, until $\vec q(t)$ and $\vec p(t)$ are themselves known. Such formulation is sufficient for deriving the equations of motion, but not well suited for canonical perturbation theory, action-angle methods, or Hamilton-Jacobi techniques. 
By contrast, expressing the system in terms of a double-variable Hamiltonian (which is fully symplectic and autonomous), standard Hamiltonian tools are mostly applicable. Thus, the already-canonical conservative sector can be appended with nonconservative radiation-reaction terms, both treated on the same footing.

For gravitational wave applications, the resulting solutions may be useful because they separate the secular inspiral from the orbital-timescale corrections within a single canonical construction. The two can be combined in a controlled manner to produce analytical solutions which balance precision and computational efficiency. This is particularly relevant for long-duration signals where small phase errors accumulate over many cycles.
Although compact binaries provide the main application of this work, the construction is not specific to gravity. Beyond compact binaries, the same strategy can be applied to any suitable perturbed integrable system. This includes systems where dissipation is initially given as a force law or reconstructed from flux-balance arguments, and not necessarily given through a Hamiltonian.

The paper is organized as follows. 
In Sec.\ \ref{sec:recall}, we recall the necessary ingredients of the doubled Hamiltonian framework and of the ordinary Lie-series method. 
in Sec.~\ref{sec:nonconservative_lie_series}, we formulate the Lie series in a nonconservative setting, leading to an analogue of the homological equation in double variables. A direct solution is obstructed by the doubled set of frequencies and resonances that appear near the physical limit. We demonstrate how this obstruction can be bypassed by introducing a weak form of the homological equation which reduces the problem back to the conservative case. 
In Sec.~\ref{sec:pn_dynamics}, we apply the construction to the leading radiation-reaction dynamics of a non-spinning compact binary in post-Newtonian regime. We start from the $2.5$PN ADM radiation-reaction forces and reconstruct the corresponding antisymmetric Hamiltonian in the doubled phase space. We then express the perturbation in osculating Delaunay variables associated with the Newtonian Kepler problem and solve the weak homological equations. The resulting resonant normal form encodes the secular inspiral and reproduces the Peters-Mathews equations \cite{Peters1963} for the semi-major axis and eccentricity. The generator then supplies the complementary oscillatory corrections, giving the full time-domain solution at leading dissipative order. This construction accommodates eccentric orbits without using a small-eccentricity expansion.

\section{The double-variable framework and the Lie-series method}
\label{sec:recall}

\subsection{Notation and properties of the double-variable structure}

We now collect the notation and the structural properties from the doubled Hamiltonian framework derived in {\hampaper} that will be used throughout the paper. The purpose of this subsection is to make clear how the doubled variables are manipulated, how the physical dynamics is extracted, and which part of the extended Hamiltonian is physically relevant.

We adopt the notation of {\hampaper}. 
Bold italic symbols (e.g.\ $\vec u$) denote vectors of $\mathbb R^N$ and bold upright symbols (e.g., $\mat M$) denote linear maps; $\tr{\mat M}$ denotes the transpose. Vectors can be split into magnitude $u = \norm{\vec u}$ and direction $\uvec u = \vec u / u$. $\vec u \cdot \vec v$ denotes the Euclidean canonical dot product. We do not distinguish between covariant and contravariant forms; namely, partial derivatives acting on scalars are interpreted as vectors, and partial derivatives acting on vectors are interpreted as linear forms:
\begin{equation}
    \mat M = \frac{\partial \vec u}{\partial \vec x} \implies M_{ij} = \frac{\partial u_j}{\partial x_i} \text,
\end{equation}
so that partial derivatives are by convention applied on the left.
We reserve the latin indices $i, j,$ and $k$ to refer to vector components, with Einstein summation applied over repeated indices. 
Perturbative order in any bookkeeping parameter is indicated with a superscript; for instance, $\hamstd^\ell$ is the $\ell$-th coefficient of the expansion of Hamiltonian $\hamstd$.

The doubled phase space consists of two labeled copies of the ordinary phase-space variables,
\begin{equation}
    (\vec q,\vec \pi) \quad \longrightarrow \quad
    (\pathA{\vec q},\pathA{\vec \pi};\, \pathB{\vec q},\pathB{\vec \pi}) \text,
\end{equation}
which are allowed to vary independently off-shell so that a stationary-action principle can encode the initial-value problem. For any single-phase-space scalar $f = f(\vec q,\vec \pi, t)$, we shall use the $(\labelA, \labelB)$ labels to denote a function's evaluation in each doubled phase space variable:
\begin{align}
    \pathA{f} &= f(\pathA{\vec q},\pathA{\vec \pi},t) \text, &
    \pathB{f} &= f(\pathB{\vec q},\pathB{\vec \pi},t) \text,
\end{align}
and introduce the $(+, -)$ or Keldysh combinations
\begin{align}
    f_+ &= \frac{1}{2}(\pathA{f}+\pathB{f}) \text, &
    f_- &= \pathA{f}-\pathB{f} \text,
\end{align}
representing the path average and path difference, respectively. We shall collectively denote $\pathAB{f} = (\pathA{f}, \pathB{f})$ and $f_\pm = (f_+, f_-)$.
For the fundamental coordinates and momenta this gives
\begin{align}
    \vec q_+ &= \frac{\pathA{\vec q}+\pathB{\vec q}}{2} \text, &
    \vec q_- &= \pathA{\vec q}-\pathB{\vec q} \text, \\
    \vec \pi_+ &= \frac{\pathA{\vec \pi}+\pathB{\vec \pi}}{2} \text, &
    \vec \pi_- &= \pathA{\vec \pi}-\pathB{\vec \pi} \text.
\end{align}
We refer to $\vec q_-$ and $\vec \pi_-$ as the virtual variables. They measure the displacement away from the physical slice of the doubled phase space, and play the role of virtual displacements in the action. Meanwhile, the average variables $\vec q_+$ and $\vec \pi_+$ collapse into the ordinary physical variables after the equations of motion have been obtained, a procedure which is called the \emph{physical limit}.

In brief, the physical limit is the operation that identifies the two labelled paths $\pathA{\vec q} = \pathB{\vec q} = \vec q$ and momenta $\pathA{\vec \pi} = \pathB{\vec \pi} = \vec \pi$ after the variational or Hamiltonian equations have been formed. 
In {\hampaper}, we demonstrated that the unique physical solution to the Hamiltonian dynamics is given precisely through the application of the physical limit condition to Hamilton's equations. This operation is represented by
\begin{equation}
    [F(\pathA{\vec q},\pathB{\vec q}, \pathA{\vec \pi}, \pathB{\vec \pi})]_{\PL} = F(\vec q, \vec q, \vec \pi, \vec \pi)
    \text,
\end{equation}
or equivalently, $[F(\vec q_+,\vec q_-,\vec \pi_+,\vec \pi_-)]_{\PL} = F(\vec q,\vec 0,\vec \pi,\vec 0)$.
The submanifold in which the physical limit condition is verified is called the \emph{physical slice}. 
A useful rule of thumb is that the physical limit should always be applied only \emph{after} the equations of motion have been produced. If one sets $\vec q_- = \vec\pi_- = \vec 0$ too early, the nonconservative information carried by the virtual derivatives would be erased. This is the Hamiltonian analogue of retaining a virtual displacement while deriving virtual-work equations, and setting that displacement to zero only at the end.

On the ordinary physical phase space we use the standard Poisson bracket
\begin{equation}
      \{ f, g \} = \frac{\partial f}{\partial \vec q} \cdot \frac{\partial g}{\partial \vec \pi} - \frac{\partial f}{\partial \vec \pi} \cdot \frac{\partial g}{\partial \vec q} \text.
\end{equation}
On the doubled phase space, the corresponding extended bracket is
\begin{equation}
     \lPB f, g \rPB = \pathmetric \left( \frac{\partial f}{\partial \vec q_a} \cdot \frac{\partial g}{\partial \vec \pi_b} - \frac{\partial f}{\partial \vec \pi_a} \cdot \frac{\partial g}{\partial \vec q_b} \right) \text,
\end{equation}
where $\pathmetric[ab]$ represents the path-label metric, and the latin subscript indices are summed through $a, b \in \{ \labelA, \labelB \}$ or $a, b \in \{ +, - \}$, which indicates the chosen labeling of the doubled paths. In the standard $(\labelA, \labelB)$ parametrization, the metric coefficients are $[\pathmetric] = \diag(1, -1)$; in the $(+, -)$ variable set, the coefficients are $[\pathmetric] = \antidiag(1, 1)$. The extended bracket is bilinear and antisymmetric, and obeys the Leibniz and Jacobi identities. In particular, the following algebra can be easily derived and will be used repeatedly:\!
\begin{subequations}\label{eq:pm_vars_bracket_algebra}
\begin{align}
    \lPB f_+, g_- \rPB & = \{ f, g \}_+  \text, \\
    \lPB f_-, g_- \rPB & = \{ f, g \}_- \text, \\
    \lPB f_+, g_+ \rPB & = \{ f, g \}_- / 4 \text.
\end{align}
\end{subequations}
The first identity is especially important. A bracket between a physical observable lifted as $f_+$ and a Hamiltonian term linear in a virtual variable produces an ordinary physical Poisson bracket.

For a system with extended Hamiltonian $\hamext = \hamext(\vec q_\pm,\vec \pi_\pm,t)$, the physical equation for an ordinary observable $f = f(\vec q,\vec \pi)$ is obtained from
\begin{equation}
    \dot f = \lPB f_+,\hamext\rPB_{\PL} \text. \label{eq:physical_equation_from_A}
\end{equation}
As in \citet{Aykroyd2024}, when dealing with the Lie series, we shall distinguish an observable evolving according to the dynamics of the Hamiltonian $\mathcal A$ from an observable evolving according to the normal-form $\mathcal A^*$. This distinction indicates respectively the instantaneous (i.e., complete with oscillatory corrections) motion and the secular (i.e., averaged) motion, which we shall denote via the tilde and overbar, namely
\begin{align}
    \dot{\tilde f} &=  \lPB {f}_+, \hamext \rPB _\mathrm{PL} \text, &
    \dot{\bar f} &=  \lPB {f}_+, \hamext^* \rPB _\mathrm{PL} \text.\label{eq:ham_eqs_single_phase_space}
\end{align}
Further details will be provided in Sec.\ \ref{sec:lie_series_guide}.

A fundamental property of the double-variable framework is that the extended Hamiltonian be antisymmetric under label exchange,
\begin{equation}
    \hamext(\pathA{\vec q},\pathB{\vec q},\pathA{\vec \pi},\pathB{\vec \pi})
    =-\hamext(\pathB{\vec q},\pathA{\vec q},\pathB{\vec \pi},\pathA{\vec \pi}) \text.
\end{equation}
This condition is the Hamiltonian counterpart of the label-exchange antisymmetry of the doubled action in \hampaper. It implies that $\hamext$ itself vanishes on the physical slice, so that its numerical value is not a relevant quantity. Instead, the derivatives of $\hamext$ transverse to the physical slice hold the information about the dynamics.
This observation motivates notation for applying Taylor expansions around the physical slice. We define the following Landau asymptotic notation, where $\vorder{k}$ denote terms of total virtual order at least $k$. Namely,
\begin{equation}
    \vorder{n}= \sum_{m = 0}^n \mathcal O\!\big(\lVert\vec q_-\rVert^m \lVert\vec \pi_-\rVert^{n-m}\big) \text,
\end{equation}
Each Landau symbol in the sum carries its own dimensionful constant.
Near the physical slice, any sufficiently regular label-antisymmetric Hamiltonian admits an odd Taylor expansion in the virtual variables. In particular,
\begin{equation}
    \hamext(\vec q_\pm,\vec \pi_\pm)=
    \vec q_-\cdot \hamext_{\vec q}(\vec q_+,\vec \pi_+)
    +\vec \pi_-\cdot \hamext_{\vec \pi}(\vec q_+,\vec \pi_+)
    + \vorder{3} \text,
\end{equation}
where $\hamext_{\vec q}$ and $\hamext_{\vec \pi}$ are coefficients of the Taylor expansion.
Only the terms linear in $\vec q_-$ and $\vec \pi_-$ contribute to Eq.~\eqref{eq:physical_equation_from_A}. Differentiating a term of order $\vorder{3}$ with respect to a virtual variable still leaves a term at least quadratic in the virtual variables, which vanishes under $[\argdot]_{\PL}$. This is why, later in the paper, two extended Hamiltonians that agree up to first virtual order are treated as physically equivalent even if they differ by $\vorder{3}$ terms.

\subsection{A brief guide to the Lie-series method}\label{sec:lie_series_guide}

Before introducing the nonconservative extension, it is useful to recall what is meant by a Lie-series perturbation method in the ordinary Hamiltonian setting. The method is a way of replacing a difficult Hamiltonian by a simpler, canonically equivalent one, order by order in a small parameter $\varepsilon$, which in our application will be the PN bookkeeping parameter. The general objective is to search for a near-identity canonical transformation that simplifies the Hamiltonian by removing certain non-essential oscillatory terms. This transformed Hamiltonian is referred to as the normal-form. Once the dynamics generated by the normal-form has been solved, the same canonical transformation can be used, perturbatively, to reconstruct the dynamics in the original variables.

The starting point is a quasi-integrable Hamiltonian
\begin{equation}\label{eq:quasi_integrable_guide}
    \mathcal H = \mathcal H^0 + \sum_{\ell\geq 1}\varepsilon^\ell \mathcal H^\ell \text,
\end{equation}
written as the sum of an integrable zeroth-order part $\mathcal H^0$, and perturbative contributions $\varepsilon^\ell \mathcal H^\ell$ for $\ell \geq 1$. The objective is to simplify the dynamics of $\mathcal H$ by transferring, as far as possible, the non-essential oscillatory dependence of the perturbation into a canonical change of variables.

For this, we consider $g = g(\vec q,\vec p)$ a smooth scalar function of phase-space, called the generator. The Poisson bracket between the generator and other phase-space functions defines a derivation,
\begin{equation}\label{eq:lie_operator_guide}
    \mathfrak L_g (f) = \{f,g\} \text.
\end{equation}
Exponentiating this derivation gives rise to the Lie series
\begin{equation}\label{eq:lie_transform_guide}
    \mathcal T_g (f) = e^{\mathfrak L_g} f
    = f + \{f,g\} + \frac{1}{2}\big\{\{f,g\},g\big\} + \cdots \text.
\end{equation}
The transformation $\mathcal T_g$ is said to be ``near-identity'' when $g$ itself is perturbatively small. In that case, the generator can be expanded as $g=\sum_{\ell\geq 1}\varepsilon^\ell g^\ell$, so that $\mathcal T_g (f)$ differ from $f$ only through controlled perturbative corrections.

With the convention used in this paper, new transformed variables $(\vec q^*, \vec p^*)$ are defined implicitly via
\begin{align}\label{eq:coordinate_lie_transform_guide}
    \vec q = \mathcal T_g(\vec q^{*}) \text, &&
    \vec p = \mathcal T_g(\vec p^{*}) \text,
\end{align}
and the inverse map is obtained by changing the sign of the generator,
\begin{equation}
    \mathcal T_g^{-1}=\mathcal T_{-g} \, .
\end{equation}
Since $\mathfrak L_g$ is the infinitesimal action of a Hamiltonian flow, $\mathcal T_g$ is canonical, preserving both the Poisson bracket and the Hamiltonian form of the equations of motion.

Equations \eqref{eq:lie_transform_guide} and \eqref{eq:coordinate_lie_transform_guide} show that a large class of near-identity canonical transformations can be parametrized by the choice of the generator $g$. The perturbative problem is therefore reduced to the question: which generator produces a transformed Hamiltonian with the simplest useful structure?

The transformed Hamiltonian is defined by preserving its value under the canonical change of variables,
\begin{equation}\label{eq:normal_form_transform_guide}
    \mathcal H^*(\vec q^*,\vec p^*)
    = \mathcal T_g(\mathcal H)(\vec q^*,\vec p^*) \text.
\end{equation}
Expanding Eq.~\eqref{eq:normal_form_transform_guide} in powers of $\varepsilon$ yields, at each order, an equation of the form
\begin{equation}\label{eq:homological_guide}
    \{g^\ell,\mathcal H^0\} = \mathcal P^\ell-\mathcal H^{*,\ell}\text,
\end{equation}
called a \emph{homological equation}. Here, $\mathcal P^\ell$ is known from the original Hamiltonian and the lower-order generators. The two unknowns are the normal-form term $\mathcal H^{*,\ell}$ and the generator term $g^\ell$. The standard choice is to set the normal-form to the \emph{resonant} part of $\mathcal P^\ell$ that cannot (or should not) be removed from the dynamics. Typically, this is a function of action variables only, and can be found via
\begin{equation}\label{eq:average_normal_form_guide}
    \mathcal H^{*,\ell}=\left\langle \mathcal P^\ell\right\rangle \text, \qquad
    \{g^\ell,\mathcal H^0\}=\mathcal P^\ell-\left\langle \mathcal P^\ell\right\rangle \text,
\end{equation}
where $\langle\argdot\rangle$ denotes the average along the unperturbed orbit described by $\mathcal H^0$. In action-angle variables, the operator $\langle\argdot\rangle$ is simply the average over the fast angles. In the two-body problem, this corresponds to an orbital average along the Keplerian ellipse. Solving the second part of Eq.\ \eqref{eq:average_normal_form_guide} for $g^\ell$ then produces a primitive (i.e., antiderivative) which encodes the remaining oscillatory part of the Hamiltonian.
This gives a useful interpretation of the construction. The normal-form Hamiltonian governs the slow, secular or averaged evolution, while the generator stores the short-period corrections required to return from the averaged variables to the original variables. 

The practical procedure is as follows. First, one solves the homological equation \eqref{eq:homological_guide} order by order, thereby determining both the normal-form Hamiltonian and the generator. Second, one solves the dynamics generated by $\mathcal H^*$, which can typically be done trivially in action-angle variables. Third, one applies the Lie map through Eq.\ \eqref{eq:coordinate_lie_transform_guide}, to recover the original variables and hence the full perturbative motion, including orbital-timescale corrections.
This is the sense in which the Lie formulation supplies both the long-term averaged evolution and the short-period corrections within a single canonical framework. In the conservative PN two-body problem, this strategy was used in our previous work \citet{Aykroyd2024} to obtain both the normal form and the associated full time-domain mapping of the conservative motion at $2$PN order. The present paper follows the same logic, but replaces the ordinary Hamiltonian and Poisson bracket by their doubled, nonconservative counterparts.
 
\section{Nonconservative Lie series}
\label{sec:nonconservative_lie_series}

\subsection{Nonconservative Hamiltonian}

We seek to generalize the standard Lie-series perturbation method to nonconservative systems. For this we will heavily employ the doubled phase-space formulation developed in {\hampaper}, which allows a Hamiltonian treatment of nonconservative processes. We begin from an autonomous extended Hamiltonian in perturbative form:
\begin{align}
\mathcal A(\pathAB{\vec q}, \pathAB{\vec \pi}) & = A^{0}(\pathAB{\vec q}, \pathAB{\vec \pi})\nonumber\\
& + \sum_{\ell=1}^{K-1} \varepsilon^\ell \mathcal A^{\ell}(\pathAB{\vec q}, \pathAB{\vec \pi}) + \mathcal O(\varepsilon^K) \text,
\end{align} 
where the leading order term is assumed integrable and conservative so that it can be separated in the doubled phase-space as $\mathcal A^0 = \mathcal H^0_- \equiv \pathA{\mathcal H}^0 - \pathB{\mathcal H}^0$. Accordingly, the nonconservative contributions are assumed to enter through some orders $\ell \geq 1$.

In practice, several distinct routes lead to an explicit construction of $\mathcal A$. 
Whenever a variational principle is available in the form of a nonconservative Lagrangian $\Lambda$ in doubled configuration space $(\pathAB{\vec q}, \pathAB{\dot{\vec q}})$, the most direct construction is via a perturbative expansion of the Legendre transform with respect to the small parameter $\varepsilon$.
In the absence of such convenience, one may start from a conservative Hamiltonian $\mathcal H$, double the phase-space variables, and then introduce a nonconservative sector $\mathcal R$. In this paradigm, the full nonconservative Hamiltonian reads:
\begin{equation} \label{eq:ham_diss_dec}
    \mathcal A = \mathcal H_- + \mathcal R \text,
\end{equation}
where the coupling term $\mathcal R$ can be generically decomposed as a linear function of the virtual $(-)$ phase-space variables
\begin{align} \label{eq:ham_diss_dec_R}
    \mathcal R(\vec q_\pm, \vec \pi_\pm) & = \vec q_- \cdot \mathcal R_{\vec{q}}(\vec q_+, \vec \pi_+) \nonumber\\
    & + \vec \pi_- \cdot \mathcal R_{\vec{\pi}}(\vec q_+, \vec \pi_+) + \vorder{3} \text.
\end{align}
When the underlying equations of motion are known, one may easily ``reverse engineer’’ the coefficients $\mathcal R_{\vec q}$ and $\mathcal R_{\vec \pi}$ \cite[see][]{Aykroyd2025}. Alternatively, one may derive $\mathcal R$ by modelling the system coupled to an environment, doubling the degrees of freedom, then integrating out the environmental variables. To see how Eq.~\eqref{eq:ham_diss_dec_R} encodes a force law, insert $\mathcal A=\mathcal H_-+\mathcal R$ into the physical Hamilton equations. Up to terms that vanish in the physical limit, one obtains
\begin{subequations}
\begin{empheq}[left=\empheqlbrace]{align}
    \dot{\vec q} &= \frac{\partial \mathcal H}{\partial \vec\pi} + \mathcal R_{\vec\pi}\text,\\
    \dot{\vec\pi} &= -\frac{\partial \mathcal H}{\partial \vec q} - \mathcal R_{\vec q}\text.
\end{empheq}
\end{subequations}
Thus $\mathcal R_{\vec\pi}$ and $-\mathcal R_{\vec q}$ are exactly the nonconservative additions to the velocity and momentum equations. This also explains why higher odd powers such as $\vorder{3}$ can be safely discarded from the Hamiltonian; their contributions vanish after taking the physical limit.

\subsection{Normal form, canonical transformation, and homological equation}

As with the classical Lie series approach, our ultimate goal is to construct a canonical transformation that carries the original Hamiltonian $\mathcal A$ into a normal form:
\begin{align}
\mathcal A^*(\pathAB{\vec \theta}^*, \pathAB{\vec K}^*) 
& = \mathcal A^{*,0}(-,\pathAB{\vec K}^*) \nonumber\\
& + \sum_{\ell=1}^{K-1} \varepsilon^\ell \mathcal A^{*,\ell}(-,\pathAB{\vec K}^*) + \mathcal O(\varepsilon^K) \text,
\end{align}
depending only on the action variables of the unperturbed motion $\pathAB{\vec K}^*$.
However, for nonconservative systems, the above form is generally not possible due to resonances.
Specifically, when doubling the degrees of freedom, we effectively introduce a new set of angles $\pathA{\theta}$ and $\pathB{\theta}$ with identical frequencies in the physical limit---the regime in which the doubled paths coincide. Thus, in a neighbourhood of the physical limit, these frequencies become resonant, so that the minus combination $\vec \theta_- = \pathA{\theta} - \pathB{\theta}$ has zero frequency, and cannot be completely removed from the normal form. In practice, these resonances will eventually lead to a dissipative secular evolution of the actions $\vec K_+^*$. As such, we must content ourselves with isolating the resonances:
\begin{align}
\mathcal A^*(\pathAB{\vec \theta}^*, \pathAB{\vec K}^*) 
& = \mathcal A^{*,0}(-,\pathAB{\vec K}^*) + \sum_{\ell=1}^{K-1} \varepsilon^\ell \Big\{ \mathcal A^{*,\ell}_\mathrm{NRES}(-,\pathAB{\vec K}^*)\nonumber\\
& + \mathcal A^{*,\ell}_\mathrm{RES}(\vec \theta^*_-, \pathAB{\vec K}^*) \Big\} + \mathcal O(\varepsilon^K) \text, \label{eq:ham_normal_form_res}
\end{align}
where as we shall see, $A^{*,\ell}_\mathrm{RES}$ will be a linear function of the resonant angles $\vec \theta^*_-$ (or equivalently of their sines).

To reach this goal [Eq.\ \eqref{eq:ham_normal_form_res}], we consider the Lie group of near-identity canonical transformations, parametrized by the action of smooth generators
\begin{equation}
    \mathfrak g(\pathAB{\vec q}, \pathAB{\vec \pi}) = \sum_{\ell=1}^{K-1} \varepsilon^{\ell} \mathfrak g^{\ell}(\pathAB{\vec q}, \pathAB{\vec \pi}) + \mathcal O(\varepsilon^K) \text,
\end{equation}
and define the Lie series via
\begin{subequations}
  \begin{empheq}[left=\empheqlbrace]{align}
    \vec q_a & = \mathcal T_{\mathfrak g}(\vec q_a^*)\text,\\
    \vec \pi_a &= \mathcal T_{\mathfrak g}(\vec \pi_a^*)\text, 
    \label{eq:phase_space_transformation}
  \end{empheq}
\end{subequations}
for $a \in \{\labelA, \labelB\}$.
As usual, we impose that the Lie transform preserves the value of the Hamiltonian for every combination of doubled phase-space variables\footnote{As we shall see, this condition is overly restrictive for practical purposes, and shall be relaxed in the subsequent section.}. In other words, we have
\begin{equation}
    \mathcal A^*(\pathAB{\vec q}^*, \pathAB{\vec \pi}^*) = \mathcal A(\pathAB{\vec q}, \pathAB{\vec \pi}) = \mathcal T_{\mathfrak g} (\mathcal A)(\pathAB{\vec q}^*, \pathAB{\vec \pi}^*) \text. \label{eq:hom_ham_lie}
\end{equation}
Expanding the Lie operator, inserting the perturbed expressions of $\mathcal A^*$ and $\mathfrak g$, and collecting order-by-order allows us to identify the double-variable homological equations:
\begin{equation}
    \lPB \mathfrak g^\ell, \mathcal A^0 \rPB = \mathcal W^\ell - \mathcal A^{*,\ell} \text, \quad \ell \in \mathbb N \text, \label{eq:hom_diss}
\end{equation}
where the perturbations $\mathcal W^\ell$ are expressions depending on the solutions of previous perturbative orders. 

Whenever the perturbation at order $\ell$ is \emph{conservative}---that is, separable with respect to the two paths $\mathcal W^\ell = \mathcal P^\ell_- = \pathA{\mathcal P}^\ell - \pathB{\mathcal P}^\ell$---the solution to the nonconservative homological equation can be directly obtained from the solution to the simpler conservative homological equation:
\begin{equation}
    \{ g^\ell, \mathcal H^0 \} = \mathcal P^\ell - \mathcal H^{*,\ell} \text, 
\end{equation}
whose unknowns are the conservative generator $g^\ell$ and normal-form Hamiltonian $\mathcal H^{*,\ell}$. Then, the double-variable homological equations [Eq.\ \eqref{eq:hom_diss}] are solved by taking $\mathcal A^{*,\ell} = \mathcal H^{*,\ell}_- = \pathA{\mathcal H^{*,\ell}} - \pathB{\mathcal H^{*,\ell}}$ and $\mathfrak g^\ell = g^\ell_- = \pathA{g^\ell} - \pathB{g^\ell}$.

For the nonconservative orders, however, there is no easy rescue. 
In classical mechanics, there exists a formal, standard solution to Eq.\ \eqref{eq:hom_diss} expressing the normal-form $\mathcal A^{*,\ell}$ as an average along the doubled flow of $\mathcal A^0 = \mathcal H^0_-$ and expressing the generator $\mathfrak g^\ell$ as a primitive.
In practice, however, the analytic evaluation of such integrals is obstructed by the doubling of phase-space dimensions---the two paths must be treated independently in an open set around the physical slice. For instance, in the perturbed Kepler problem, this involves simultaneously integrating nonlinear functions evolved across two independent ellipses. Moreover, the doubling of phase-space introduces resonances inherent to the degeneracy of the frequencies as $\pathA{\vec q} \sim \pathB{\vec q}$. Because resonances cannot in general be eliminated via the Lie procedure, this presents a real obstruction for solving Eq.~\eqref{eq:hom_diss} directly. An alternate resolution strategy is therefore required.

\subsection{Weak homological equation and solutions}


To circumvent the obstacle, we recall an essential structural property arising in the double-variable framework: two Hamiltonians are physically equivalent if their first-order Taylor expansions in the virtual variables $(\vec q_-, \vec \pi_-)$ coincide \cite[see][]{Aykroyd2025}. Thus, terms of order $\vorder{3}$ never contribute once the physical limit is enforced.
This equivalence allows one to redefine the normal-form Hamiltonian $\mathcal A^*$ by an arbitrary $\vorder{3}$ shift in Eq.\ \eqref{eq:hom_ham_lie} without affecting the dynamics\footnote{In truth, we are redefining the dynamics far away from the physical slice while keeping the physical dynamics intact.}, which gives rise to a weak version of the nonconservative homological equation:
\begin{equation}
    \lPB \mathfrak g^\ell, \mathcal A^0 \rPB = \mathcal W^\ell - \mathcal A^{*, \ell} + \vorder{3} \text. \label{eq:hom_diss_weak}
\end{equation}

The functions involved in this equation are all anti-symmetric with respect to $(\labelA, \labelB)$ label exchange, which means that $f(\pathA{\vec q}, \pathB{\vec q}, \pathA{\vec \pi}, \pathB{\vec \pi}) = - f(\pathB{\vec q}, \pathA{\vec q}, \pathB{\vec \pi}, \pathA{\vec \pi})$. Equivalently, these functions have an odd parity in the virtual variables $(\vec q_-, \vec \pi_-)$ when Taylor-expanded around the physical limit. Accordingly, for an arbitrary parameter set $\vec x = \vec x(\vec q, \vec \pi)$ forming a diffeomorphism in phase-space, we decompose
\begin{equation}
    f(\vec q_\pm, \vec \pi_\pm) = \vec x_- \cdot f_{\vec x}(\vec q_+, \vec \pi_+) + \vorder{3} \text. \label{eq:asym_dec}
\end{equation}
Here, $f_{\vec x} = (f_{x^1}, \ldots, f_{x^{2N}})$ are $2N$ functions whose explicit phase-space  expressions are 
\begin{equation} \label{eq:asym_dec_coeff}
    f_{\vec x}(\vec q, \vec \pi) =
    \pars*{ 
        \frac{\partial \vec q}{\partial \vec x} \, \dext{f}{\vec q_-} 
        + \frac{\partial \vec \pi}{\partial \vec x} \, \dext{f}{\vec \pi_-}
    }_{\!\PL} \text.
\end{equation}
We emphasize that the coefficients $f_{\vec x}$ do not need to be expressed in terms of the variables $\vec x$---that is, they can be kept as functions of $\vec q_+$ and $\vec \pi_+$---and that the variable set $\vec x$ does not need to obey any canonicity properties. 

The key step is to apply the decomposition \eqref{eq:asym_dec} to all functions of Eq.\ \eqref{eq:hom_diss_weak} except for the unperturbed Hamiltonian which is kept in the form $\mathcal A^0 = \mathcal H^0_-$.
Namely, we perform the substitutions
\begin{subequations}
\begin{align}
    \mathfrak g^\ell &= \vec x_- \cdot \mathfrak g^\ell_{\vec x} + \vorder{3}, \\
    \mathcal W^\ell &= \vec x_- \cdot \mathcal W^\ell_{\vec x}  + \vorder{3}, \\ 
    \mathcal A^{*,\ell} &= \vec x_- \cdot \mathcal A^{*,\ell}_{\vec x}  + \vorder{3} .
\end{align}
\end{subequations}
We highlight that the only terms from the generator that can affect the right-hand side (RHS) at $\vorder{3}$ are those of linear virtual order [since $\lPB \vorder{3}, \mathcal H^0_- \rPB = \vorder{3}$, c.f.\ the properties of Eqs.\ \eqref{eq:pm_vars_bracket_algebra}].

Using the linearity and Leibniz rules of the doubled Poisson bracket together with the
$(+,-)$ algebraic identities [Eqs.\ \eqref{eq:pm_vars_bracket_algebra}], we distribute the Poisson brackets in Eq.\ \eqref{eq:hom_diss_weak} with the expanded forms. Then, since the homological equation is a functional equation, we can collect and equate resulting coefficients linearised in $\vec x_-$.
We are left with a system of coupled differential equations in the physical phase-space:
\begin{equation}
    \{ \mathfrak g^\ell_{\vec x}, \mathcal H^0 \} + \mat M \,  \mathfrak g^\ell_{\vec x} = \mathcal W^\ell_{\vec x} - \mathcal A^{*,\ell}_{\vec x} \text,  
    \label{eq:hom_diss_pde}
\end{equation}
Here $\{\argdot, \argdot\}$ is the single phase-space (physical) Poisson bracket. The generator coefficients $\mathfrak g^\ell_{\vec x}$ and Hamiltonian coefficients $\mathcal A^{*,\ell}_{\vec x}$ are treated as independent functions to be determined as solutions to the differential equation, whereas the perturbation coefficients $\mathcal W^\ell_{\vec{x}}$ are explicitly computed from the decomposition of $\mathcal W^\ell$ [via Eq.\ \eqref{eq:asym_dec_coeff}]. The transformation matrix $\mat M$ is constructed from the Jacobian of the unperturbed flow,
\begin{equation}
    [\mat M]_{ij} = \frac{\partial}{\partial x^i} \Big( \{ x^j, \mathcal H^0 \} \Big)
\end{equation}
and depends strictly on the choice of parameter set $\vec x$, which, at this step, is still arbitrary.

The utility of \eqref{eq:hom_diss_pde} hinges on choosing $\vec x$ so that $\mat M$ simplifies. The naive Cartesian choice $\vec x=(\vec q,\vec \pi)$ produces a nontrivial coupling and is not recommended. For instance, in the two-body problem, $\vec x = (\vec q, \vec \pi)$ yields the block form
\begin{equation}
    \mat M = \begin{pmatrix}
        0 && \mat Q \\
        \idmat{} && 0
    \end{pmatrix},
\end{equation}
where $\mat Q$ is a quadrupole term
\begin{equation}
     \mat Q = \frac{1}{q^3} \left( 3 \uvec q \otimes \uvec q - \idmat{} \right) \text.
\end{equation}
As a consequence of this choice, the resulting coupled homological equation reduces to a second order system whose solution is obscured. 

Rather, it is more advantageous to work with the virtual expansion of the action-angle variables $\vec x = (\vec \theta, \vec K)$, whose brackets are given via
\begin{subequations}
\begin{align}
    \{ \vec \theta, \mathcal H^0 \} &= \vec \omega(\vec K), \\
    \{ \vec K, \mathcal H^0 \} &= \vec 0.
\end{align}
\end{subequations}
Remark: to keep the linear expression of Eq.\ \eqref{eq:asym_dec} globally well–defined on the torus, bounded and continuous, we replace each virtual angle $\theta_-$ by the sinusoidal $s_{\theta_-} \equiv \sin \theta_- = \theta_- + \vorder{3}$, so that
\begin{align}
    f(\vec q_\pm, \vec \pi_\pm) & = \vec s_{\vec \theta_-} \cdot f_{\vec \theta}(\vec q_+, \vec \pi_+) \nonumber\\
    & + \vec K_- \cdot f_{\vec K}(\vec q_+, \vec \pi_+) + \vorder{3} \text.\label{eq:asym_dec_angles}
\end{align}
In this basis the transformation matrix
\begin{equation}
    \mat M = \begin{pmatrix}
        \mat 0 && \mat 0 \\
        \grad \vec \omega(\vec K) && \mat 0
    \end{pmatrix}
\end{equation}
is block-triangular, and the homological equations in physical space [Eq.\ \eqref{eq:hom_diss_pde}] become partially decoupled:
\begin{subequations}\label{eq:hom_diss_ode}
\begin{align}
    &\{ \mathfrak g_\theta^\ell, \mathcal H^0 \} = \mathcal W_\theta^\ell - \mathcal A_\theta^{*, \ell} \text, \label{eq:hom_diss_ode:angle}\\
    &\{ \mathfrak g_K^\ell, \mathcal H^0 \} = \Big( \mathcal W_K^\ell - \frac{\partial \omega}{\partial K} \mathfrak g_\theta^\ell \Big) - \mathcal A_K^{*, \ell} \label{eq:hom_diss_ode:action}\text,
\end{align}
\end{subequations}
for every action-angle pair $(\theta, K) \in (\vec \theta, \vec K)$. Now, each couple of equations for a given action-angle pair can be solved via standard methods---by first determining $\mathfrak g^\ell_\theta$ from Eq.\ \eqref{eq:hom_diss_ode:angle}, and then substituting into the equation \eqref{eq:hom_diss_ode:action} for $\mathfrak g^\ell_K$. 
The main advantage of Eqs.\ \eqref{eq:hom_diss_ode} over the original Eq.\ \eqref{eq:hom_diss} is that while the latter is formulated in the doubled phase-space, the former is expressed in the single, physical phase-space. The redundant double-variable frequencies which hindered the resolution of Eq.\ \eqref{eq:hom_diss} are no longer present.
In effect, we have reduced the resolution of the nonconservative homological equation to a set of $2N$ sequentially decoupled conservative-like homological equations. 


\section{Post-Newtonian dynamics}
\label{sec:pn_dynamics}

In order to apply our Lie method, we start from the ADM Hamiltonian describing the matter dynamics of a point-mass binary in the transverse-traceless (TT) gauge. We define the relative separation vector $\vec r$ in the center-of-mass frame and $\vec \pi$ its canonically conjugate momentum, both written in the standard ADM rescaled variables. In the ADM approach, after the gravitational-field variables are reduced and the leading radiative back-reaction is encoded in the matter sector, the $2.5$PN dissipative dynamics can be represented by an explicitly time-dependent Hamiltonian function \cite{Schafer1990,Kokkotas1995,Buonanno1999,Schafer2024}:
\begin{multline}
    \mathcal H(\vec r, \vec \pi, t) = \mathcal H^0(\vec r, \vec \pi) + \varepsilon^2 \mathcal H^1(\vec r, \vec \pi) + \varepsilon^4 \mathcal H^2(\vec r, \vec \pi) \\ + \varepsilon^5 \mathcal H^{2.5}(\vec r, \vec \pi, t) + \porder{6} \text, 
\end{multline}
containing both even- and odd-powered monomials in the bookkeeping parameter $\varepsilon = 1/c$. We denote by $\ell$ the power of $\varepsilon$, so that the corresponding PN order is $\ell/2$; the Hamiltonian perturbations are therefore expressed as $\varepsilon^\ell \mathcal H^{\ell/2}$ (a similar remark applies to the normal-form and generator).
We assume that the system is expressed in rescaled center-of-mass coordinates, so that the Newtonian term reads
\begin{equation}
  \mathcal H^0(\vec r, \vec \pi) = \frac{\pi^2}{2} - \frac{1}{r} \, .
\end{equation}
The $1$PN and $2$PN perturbations are known from the literature and we recall them in Appendix \ref{appendix:Hamiltonian_ADM}.
The leading dissipative contribution $\mathcal H^{2.5}$ arises at 2.5PN from the radiation-reaction and reads explicitly \cite{Schafer1990, Kokkotas1995, Buonanno1999, Schafer2024}:
\begin{equation}
    \mathcal{H}^{2.5}(\vec r, \vec \pi, t) = \frac{4 \nu}{5} \pars*{ 
        \frac{\pi^i \pi^j}{2} - \frac{r^i r^j}{2 r^3} 
    } \dddot{I}_{\!\!ij}(t)  \text,
\end{equation}
with $\nu = m_1 m_2 / (m_1+m_2)^2$ the symmetric mass ratio, $I_{ij} = r^i r^j - (1/3) r^2 \delta_{ij}$ the mass quadrupole moment and $\delta_{ij}$ the Kronecker delta.
In practice one treats $\dddot I_{\!\!ij}(t)$ as a function of time when computing Poisson brackets, and only once Hamilton's equations are deduced is one allowed to identify $\dddot I_{\!\!ij}(t) = \dddot I_{\!\!ij}(\vec r, \vec \pi)$, where to leading order we impose the Keplerian motion, $\dot{\vec \pi} = - \uvec r / r^2$.
This results in the following radiation-reaction contributions:
\begin{subequations}    
\label{eq:2.5PN_eom_contributions}
\begin{empheq}[left=\empheqlbrace]{align}
    \Delta \dot{\vec \pi} &= - \varepsilon^5 \frac{\partial \mathcal{H}^{2.5}}{\partial \vec r} = \frac{8 \nu}{15 c^5} \frac{\left( 7 \pi_r \uvec r - 6 \vec \pi \right)}{r^4}  \text, \\
    \Delta \dot{\vec r} & = \varepsilon^5 \frac{\partial \mathcal{H}^{2.5}}{\partial \vec \pi} = \frac{8 \nu}{15 c^5} \frac{\left[ (9 \pi_r^2 - 6 \pi^2) \uvec r - 5 \pi_r \vec \pi \right]}{r^2} \text,
\end{empheq}
\end{subequations}
where $\pi_r = \pi \cdot \uvec r$.

We emphasize that $\mathcal{H}^{2.5}$ as-is is not expressed in the double-variable formalism, despite suggestions to the contrary \cite[see e.g.][]{Schafer2024}. Specifically, any double-variable Hamiltonian must be antisymmetric with respect to an interchange of the two variable labels, or else it will generate conflicting equations of motion for the two paths, which violates the physical limit property and the symplectic structure (see \hampaper).
To reconstruct a coherent Hamiltonian in doubled phase-space we instead impose the form outlined in Eqs.~\eqref{eq:ham_diss_dec} and \eqref{eq:ham_diss_dec_R}:
\begin{multline}
\label{eq:ham_ext_adm_series}
    \!\!\!\!\!
    \mathcal A(\pathAB{\vec r}, \pathAB{\vec \pi}) = \mathcal H_-^0 + \varepsilon^2 \mathcal H_-^1 + \varepsilon^4 \mathcal H_-^2 + \varepsilon^5 \mathcal R^{2.5}(\pathAB{\vec r}, \pathAB{\vec \pi}) \text.
    \!\!\!
\end{multline}
The Hamiltonian terms of the form $\varepsilon^\ell \mathcal H_-^\ell$ lead trivially to the conservative sector in Hamilton's equations, while the dissipative contributions are contained within the coupling term $\mathcal R^{2.5}$. 

The doubled Hamiltonian may be reconstructed by matching the physical vector field generated by the radiation-reaction terms to the linear virtual-variable expansion of $\mathcal R^{2.5}$.
More generally, for some time-dependent dissipative Hamiltonian of the form
\begin{equation}
    \mathcal H^\mathrm{diss}(\vec r, \vec \pi, t) = U(\vec r, \vec \pi) V(\vec r(t), \vec \pi(t)),
\end{equation}
where $\vec r(t)$ and $\vec \pi(t)$ are functions of time, solutions to the dynamics of $\mathcal H$,
a consistent Hamiltonian $\mathcal A$ in double-variables can be constructed by a \emph{label anti-symmetrization prescription}
\begin{equation}
    \mathcal A(\pathAB{\vec r}, \pathAB{\vec \pi}) = U_- V_+ \text,
\end{equation}
where the dynamical piece $U$ is replaced by the anti-symmetric part $U_-$ and the non-dynamical piece $V$ is replaced by the symmetric part $V_+$.
The resulting equations of motion can be found from the Leibniz rule and double-bracket algebra [Eq.\ \eqref{eq:pm_vars_bracket_algebra}], namely
\begin{equation}
    \dot f = \lPB f_+, U_- V_+ \rPB_\PL = \{f, U\} V\text.
\end{equation}
This equation correctly reproduces the ``dynamically inert'' feature of the time-dependent piece $V$, namely $V$ behaves as a constant inside the above bracket.
The prescription should be understood as a mapping between two distinct representations of the radiation-reaction dynamics. The standard PN representation encodes the dynamics into either a time-dependent Hamiltonian or dissipative acceleration terms. The above prescription captures that same information into the coefficients $U$ and $V$, in a manner that the resulting bracket, evaluated in the physical limit, returns the correct dissipative vector field.
This representation conversion is not unique away from the physical slice since terms of virtual order $\vorder{3}$ can be added without modifying the equations of motion.
For the $2.5$PN radiation-reaction perturbation, we directly obtain the Hamiltonian term
\begin{align}
    \mathcal R^{2.5}(\pathAB{\vec r}, \pathAB{\vec \pi}) &= \frac{4 \nu}{5} \pars*{ 
        \frac{\pi^i \pi^j}{2} - \frac{r^i r^j}{2 r^3} 
    }_{\!\!-} \pars*{\dddot{I}_{\!\!ij}}_+  \nonumber\\
    &= \frac{8 \nu}{15} \Bigg\{ 
    \frac{\left[  \pars*{ 9 {\pi_r}_+^2 - 6 \pi_+^2  } \uvec r_+ - 5 {\pi_r}_+  \vec \pi_+ \right] \cdot \vec \pi_-}{r_+^2}
    \nonumber\\
    &+\frac{\left( 6 \vec \pi_+ - 7 {\pi_r}_+\uvec{r}_+ \right) \cdot \vec{r}_-}{r_+^4}\Bigg\} \text.\label{eq:Hamiltonian_2.5PN_PMV_rp}
\end{align}
The last equality is optional, obtained from a linear expansion of $\mathcal R^{2.5}$ in the virtual $(-)$ variables [see Eq.\ \eqref{eq:asym_dec}], or alternatively, by matching the contributions to the equations of motion [Eq.\ \eqref{eq:2.5PN_eom_contributions}] through $\Delta \dot{\vec \pi} = - \varepsilon^5 \mathcal R^{2.5}_{\vec{r}}$ and $\Delta \dot{\vec r} = \varepsilon^5 \mathcal R^{2.5}_{\vec{p}}$.

Now that our double-variable Hamiltonian [Eq.\ \eqref{eq:ham_ext_adm_series}] is constructed, the next step is to re-express the perturbation $\mathcal R^{2.5}$ as a function of the virtual action-angles of the unperturbed Hamiltonian $\mathcal H_0$. We adopt the osculating Delaunay phase-space elements (see e.g., \cite{Aykroyd2024} for explicit definitions), and differ from the quasi-Keplerian parametrization \cite{Damour1985, Memmesheimer2004}, where the Newtonian relationships are adjusted with PN corrections. We construct the parameter set $\vec x = (M, \omega, \Omega, L, J, H)$, where the mean anomaly $M$, argument of the periapsis $\omega$, and longitude of the node $\Omega$ are the canonical angles associated respectively to the conjugate actions $L$, $J$ and $H$. We remark that the circular limit should be taken separately because Delaunay variables are singular there. A similar caveat applies to zero inclination orbits. Accordingly, $\mathcal R^{2.5}$ can be expanded with respect to these Delaunay variables via Eq.\ \eqref{eq:asym_dec_angles}:
\begin{align}
    \mathcal R^{2.5}(\pathAB{\vec q}, \pathAB{\vec \pi}) &= \sin M_- \mathcal R^{2.5}_M(\vec q_+, \vec \pi_+) + L_- \mathcal R^{2.5}_L(\vec q_+, \vec \pi_+) \nonumber\\ &
    + \sin \omega_- \mathcal R^{2.5}_\omega(\vec q_+, \vec \pi_+) + J_- \mathcal R^{2.5}_J(\vec q_+, \vec \pi_+) \nonumber\\ &
    + \sin \Omega_- \mathcal R^{2.5}_\Omega(\vec q_+, \vec \pi_+) + H_- \mathcal R^{2.5}_H(\vec q_+, \vec \pi_+) \text,
\end{align}
where the coefficients are obtained by applying the chain rule [Eq.\ \eqref{eq:asym_dec_coeff}], which gives
\begin{subequations}
\begin{align}
    \mathcal R^{2.5}_M(\vec r, \vec \pi) &= \frac{8 \nu }{15} \pars*{
        \frac{11 J^2 L^3}{r^6}+\frac{2 L^3}{r^5}-\frac{L}{r^4}
    } ,\\
    \mathcal R^{2.5}_L(\vec r, \vec \pi) &= \frac{8 \nu }{15}
       \Bigg[
            \frac{11 J^4}{L r^5}+\frac{1}{r^4} \left(\frac{13 J^2}{L}-\frac{9 J^4}{L^3}\right) \nonumber\\
            &+\frac{1}{r^3} \left(\frac{J^2}{L^3}+\frac{2}{L}\right)-\frac{1}{r^2} \frac{2  e^2}{L^3} 
       \Bigg] \frac{\pi_r}{e^2}
    ,\\
    \mathcal R^{2.5}_\omega(\vec r, \vec \pi) &=  \frac{8 \nu }{15} \pars*{ \frac{9 J^3}{r^5}+\frac{3 J}{L^2 r^3} } ,\\
    \mathcal R^{2.5}_\Omega(\vec r, \vec \pi) &=  \frac{8 \nu }{15} \pars*{ \frac{9 J^3}{r^5}+\frac{3 J}{L^2 r^3} } \cos\iota ,\\
    \mathcal R^{2.5}_J(\vec r, \vec \pi) &= -\frac{8 \nu }{15} \pars*{ \frac{11 J^3}{r^5}+\frac{4 J}{r^4} +\frac{3 J}{L^2 r^3}} \frac{\pi_r}{e^2} ,\\
    \mathcal R^{2.5}_H(\vec r, \vec \pi) &= 0.
\end{align}
\end{subequations}

\subsection{Lie transformation}

Up to $2$PN ($\ell \leq 4$) the ADM Hamiltonian is purely conservative, so that the normal-form and generator solutions trivially decompose into conservative and dissipative orders:
\begin{subequations}
\begin{align}
    \mathcal A^* & = \pars{ \mathcal H^* }_-^0 + \varepsilon^2 \pars{ \mathcal H^* }_-^1 + \varepsilon^4 \pars{ \mathcal H^* }_-^2 + \varepsilon^5 \pars{\mathcal R^*}^{2.5} \text, \\
    \mathfrak g & = \varepsilon^2 g_-^1 + \varepsilon^4 g_-^2 + \varepsilon^5 \mathfrak g^{2.5} \text.     \label{eq:gen_expansion}   
\end{align}    
\end{subequations}
At conservative orders $\ell = 0, 2, 4$, the corresponding functions of physical phase-space $\pars{ \mathcal H^*}^
{\ell/2}$ and $g^{\ell/2}$ are those already obtained in the literature (e.g., in \cite{Aykroyd2024}). Meanwhile, the dissipative $2.5$PN coupling term $\pars{\mathcal R^*}^{2.5}$ can be obtained from the solution to the six homological equations at $\ell = 5$ [cf.\ Eqs.\ \eqref{eq:hom_diss_ode}], with perturbation $\mathcal W^{2.5} = \mathcal R^{2.5}$. In the two-body problem, the unperturbed Hamiltonian $\mathcal H_0$ is maximally integrable, so that, in physical phase space, there is a single non-degenerate angle $M$ (the mean anomaly), with corresponding frequency $n = L^{-3} \neq 0$ (the mean motion).

We apply the integration technique of \citet{Aykroyd2024} for two-body conservative homological equations, computing the normal-form and generator.
The key computational challenge (which is absent in previous PN order calculations) is the emergence of an integral of the true anomaly $v$ over the flow of $\mathcal H_0$---which appears due to the generator term in Eq.\ \eqref{eq:hom_diss_ode:action} with $(\theta, K) = (M, L)$---and is dealt in the Appendix. We accordingly recover the normal form solution
\begin{widetext}
\begin{align} \label{eq:diss_normal_form}
    \pars{\mathcal R^*}^{2.5}(\pathAB{\vec r}, \pathAB{\vec \pi}) = \frac{\nu }{15}  
    \left(\frac{425}{J_+^7}-\frac{366}{J_+^5 L_+^2}+\frac{37}{J_+^3 L_+^4}\right) \sin M_-
    + \frac{4\nu }{5} \left(\frac{15}{J_+^4 L_+^3} - \frac{7}{J_+^2 L_+^5} \right) (\sin \omega_- + \cos \iota_+ \sin \Omega_-) 
     \text.
\end{align}
The resonant behaviour is now evident from the persistence of the ``virtual angular harmonics'' (due to $\sin M_-$, $\sin \omega_-$, and $\sin \Omega_-$) in the dissipative sector. Because these virtual angles are degenerate in the physical limit (i.e., their frequencies are equal to zero), they are resonant in the Hamiltonian and cannot be removed through the Lie series procedure. This is not a shortcoming of the method but instead indicate an intrinsic evolution of the action integrals of the secular dynamics. In contrast, the dissipative normal form above contains no virtual action terms. As a consequence, we shall see that it produces no direct $2.5$PN correction to the angle frequencies; all effects on the angles are indirect, through the dissipative evolution of the actions.

Finally, from the homological equations we likewise recover the following coefficients for the generator,
\begin{subequations}\label{eq:diss_gen}
\begin{align}
    \mathfrak g^{2.5}_M(\vec r, \vec \pi) &= -\frac{\nu }{45} \bigg\{ 
       \left[ \left( \frac{673 L}{J^4}-\frac{1275 L^3}{J^6} \right) + \left(\frac{111 L}{J^2}-\frac{425 L^3}{J^4}\right) \frac{1}{r} -\frac{170 L^3}{J^2 r^2}-\frac{66 L^3}{r^3} \right] \pi_r \nonumber \\
        & + \left(\frac{1275 L^3}{J^7}-\frac{1098 L}{J^5}+\frac{111}{J^3 L}\right) (M - v)
    \bigg\} \text,\\
    \mathfrak g^{2.5}_L(\vec r, \vec \pi) &= \frac{\nu}{15}\bigg\{ 
    \left(-\frac{22 L^3}{e^2}+22 J^2 L+22 L^3\right)\frac{1}{r^4}
    +\left(-\frac{32 L}{3 e^2}-\frac{24 J^2}{L}+\frac{32 L}{3}\right)\frac{1}{r^3}
    +\left(-\frac{12}{e^2 L}-\frac{29}{L}\right)\frac{1}{r^2} 
    +\left(\frac{16}{L^3}-\frac{170}{J^2 L}\right)\frac{1}{r}
    \nonumber\\&
    + \left(\frac{134}{3 e^2 J L^4}-\frac{3825 L^2}{J^7}-\frac{2550 L}{J^6}+\frac{5844}{J^5}+\frac{1771}{J^4 L}-\frac{2104}{J^3 L^2}-\frac{111}{J^2 L^3}+\frac{175}{J L^4}-\frac{16}{L^5}\right)\nonumber
    \\&
    +\left(-\frac{1275}{J^7}+\frac{1275}{J^6 L}+\frac{1098}{J^5 L^2}-\frac{673}{J^4 L^3}-\frac{111}{J^3 L^4}\right)r
    +\left(\frac{1275}{2 J^7 L^2}-\frac{549}{J^5 L^4}+\frac{111}{2 J^3 L^6}\right)r^2
    \nonumber\\&
    +\left( \frac{1275 L}{J^6}-\frac{673}{J^4 L} \right) \ln\pars*{\frac{2 r}{L (J+L)}}
    + \left(-\frac{1275 L}{J^7}+\frac{1098}{J^5 L}-\frac{111}{J^3 L^3}\right) \pi_r r (v - u)
    \nonumber\\&     
    +\left(-\frac{1275 L^2}{J^7}+\frac{1098}{J^5}-\frac{111}{J^3 L^2}\right) \Re\dilog\pars*{ \ee^{\ii u} \sqrt{ \frac{L - J}{L + J}}} 
    \bigg\} \text,\\
    \mathfrak g^{2.5}_\omega(\vec r, \vec \pi) &= -\frac{4 \nu }{5} \left\{ 
        \left(\frac{15}{J^4}-\frac{7}{J^2 L^2}\right) (M - v)
        -\left(\frac{15}{J^3}-\frac{2}{J L^2}+\frac{2 J}{r^2}+\frac{5}{J r}\right) \pi_r 
    \right\}
    \text, \\  
    \mathfrak g^{2.5}_J(\vec r, \vec \pi) &= -\frac{\nu}{45 e^2} \left\{ \frac{3}{L^5} + \frac{131}{J^2 L^3}-\frac{36 J}{L^2 r^2}-\frac{32 J}{r^3}-\frac{66 J^3}{r^4}\right\} \text, \\
    \mathfrak g^{2.5}_\Omega(\vec r, \vec \pi) &= \mathfrak g^{2.5}_\omega(\vec r, \vec \pi) \,\cos{\iota} ,\\
    \mathfrak g^{2.5}_H(\vec r, \vec \pi) &= 0 \text,
\end{align}
\end{subequations}
where $u$ is the eccentric anomaly.
\vspace*{-1em}
\end{widetext}

\subsection{Secular dynamics}
\label{sec:secular_dynamics}

Once the normal-form Hamiltonian $\mathcal A^*$ is known, the long–term (secular) evolution of any phase–space function $f = f(\vec r^*, \vec \pi^*)$ follows from Hamilton's equations in the physical limit,
\begin{equation}\label{eq:secular_poisson_equation}
    \frac{\dd \bar f}{\dd t} = { \lPB \bar f_+, \mathcal A^* \rPB }_\PL = {\{f, \mathcal H^{*,\mathrm{cons}} \}} + {\{ f, \vec x \} \cdot \mathcal R^*_{\vec x}} \text,
\end{equation}
where the RHS is evaluated via the standard Poisson bracket on physical phase-space, $\mathcal H^{*,\mathrm{cons}} = \sum_{\ell=0}^2 \varepsilon^{2\ell} \mathcal H^{*,\ell}$ is the conservative sector of $\mathcal A^*$, and $\mathcal R^* = \varepsilon^5 \pars{\mathcal R^*}^{2.5}$ is the dissipative sector.

Direct computation of the brackets supplies the secular motion of each of the osculating orbital elements. The two pieces in Eq.~\eqref{eq:secular_poisson_equation} play different roles in the $2.5$PN dynamics. The first, conservative term supplies the base action-angle description with PN-corrected orbital frequencies.
In contrast, the second, resonant term induces a slow evolution of the action variables, which are no longer constant in the dissipative case. In rescaled units:~
\begin{subequations}
    \begin{align}
        \!\!\!&\frac{\dd \bar a}{\dd t} = - \frac{\nu}{c^5}  \left(\!\frac{64}{5 \bar a^3 \left(1-\bar e^2\right)^{7/2}}\!\right) \! \left( \! 1 + \frac{73 \bar e^2}{24} + \frac{37 \bar e^4}{96}\right) \text,\label{eq:evo_a}\\
        \!\!\!&\frac{\dd \bar e}{\dd t} = - \frac{\nu}{c^5} \left(\frac{304 \bar e}{15 \bar a^4 \left(1-\bar e^2\right)^{5/2}}\right) \!\left(1 + \frac{121 \bar e^2}{304}\right) \text,\label{eq:evo_e}\\
        \!\!\!&\begin{aligned}
        &\frac{\dd \bar M}{\dd t} \equiv \bar n(\bar a, \bar e) \text,&  
        &\frac{\dd \bar \omega}{\dd t} \equiv \dot {\bar \omega}(\bar a, \bar e) \text,
        \end{aligned}\\
        \!\!\!&\begin{aligned}
        &\frac{\dd \bar \iota}{\dd t} = 0 \text,&
        &\frac{\dd \bar \Omega}{\dd t} = 0 \text.
        \end{aligned}
    \end{align}
\end{subequations}
In the first two equations, we recover the standard Peters-Mathews equations for the average evolution of the semi-major axis and eccentricity \cite[c.f.][]{Peters1963, Poisson2014}.
The dissipative terms produce no corrections to the angle frequencies. The functions $\bar n = \bar n(\bar a, \bar e)$ and $\dot {\bar \omega} = \dot {\bar \omega}(\bar a, \bar e)$ are the secular mean motion and periapsis frequency respectively; they originate entirely from the conservative $1$PN and $2$PN dynamics. Their explicit expressions are recalled in Appendix~\ref{app:secular_conservative_dynamics} \cite[see][]{Aykroyd2024}. The inclination $\iota$ and longitude of the node $\Omega$ are constant to the order considered here; the leading radiation reaction effect preserves the direction of the orbital angular momentum on secular timescales.

The dissipative evolution can be expressed in terms of the time solution for the eccentricity $\bar e = \bar e(t)$, which unfortunately has no elementary closed-form (unperturbed) analytical expression in terms of elementary functions\footnote{For a series solution, see \cite{Pierro1996}.}. 
Upon taking the ratio of Eqs.~\eqref{eq:evo_a} by \eqref{eq:evo_e} and integrating, we recover the well-known (exact) relation \cite[see e.g.][]{Poisson2014}
\begin{equation}
    \bar a(t)= \bar a_0 \left( \frac{1-\bar e_0^2}{1-\bar e(t)^2\!} \right) \left(\frac{\bar e(t)}{\bar e_0}\right)^{\!\!\frac{12}{19}} \! \left(\frac{\bar e(t)^2+\frac{304}{121}}{\bar e_0^2+\frac{304}{121}}\right)^{\!\!\frac{870}{2299}} \!\!\text, \label{eq:a_e_sol}
\end{equation}
where $a_0$ and $e_0$ denote the initial conditions.
Reinjecting Eq.\ \eqref{eq:a_e_sol} into the equation for $\dot{\bar e}(t)$ we recover a decoupled ordinary differential equation for the eccentricity \cite[see e.g.][]{Poisson2014}:
\begin{equation}
    \frac{\dd \bar e}{\dd t} = - \frac{\nu}{15 c^5} \frac{e_0^{48/19} \left(1-\bar e^2\right)^{3/2} \left(121 e_0^2+304\right)^{3480/2299} }{a_0^4 \bar e^{29/19} \pars*{e_0^2-1}^4 \pars*{121 \bar e^2+304}^{1181/2299} } \text.
\end{equation}
Although this equation does not to admit a closed-form integral, the problem is still tractable perturbatively, order by order in $1/c$. Since the aim is to obtain a PN-expanded solution to the dynamics, rather than an exact non-perturbative one, this perturbative treatment is sufficient.
Furthermore, notice that the derivative is singular for $\bar e \to 0$ and $\bar e \to 1$. The former reflects the simultaneous shrinkage of the semimajor axis $\bar a(\bar e) \to 0$ in the late inspiral as $\bar e \to 0$. In general, due to gravitational wave circularizing of orbits, driving the eccentricity downwards during the inspiral \cite{Peters1963, Poisson2014}, we require the PN solution to be regular at least near $\bar e \to 0$. Indeed, we find the following fractional-power expansion, truncated at leading $2.5$PN order, which is valid for $\bar e_0 \neq 1$:
\begin{equation}
\label{eq:sol_e}
    \left[ \frac{\bar e(t)}{\bar e_0} \right]^{48/19} = 1 - \frac{16 \nu}{95 c^5}\frac{\left(121 e_0^2+304\right)}{ a_0^4 \left(1-e_0^2\right)^{5/2}} \, t \text.
\end{equation}
This expression can then be reinjected into Eq.\ \eqref{eq:a_e_sol} to produce an explicit time evolution for the semimajor axis.

Once $\bar a = \bar a(t)$ and $\bar e = \bar e(t)$ are known, the osculating mean anomaly can be integrated simply through
\begin{equation}
    \bar M(t) = \bar M_0 + \int_0^t \bar n\big(\bar a(t'), \bar e(t')\big) \, \dd t' \text.
\end{equation}
The mean motion $\bar n$ contains contributions from conservative orders $0$PN, $1$PN, and $2$PN, while the time dependence of $\bar a$ and $\bar e$ begin only at $2.5$PN. We can leverage this to construct a Taylor series for $\bar n = \bar n(t)$ of the form
\begin{equation}
    \bar n(t) = \bar n_0 + \dot{\bar n}_0 t + \frac{1}{2}\ddot{\bar n}_0 t^2 + \ldots
\end{equation}
We shall see that the monomials accompanying $t^2$ and higher powers occur formally beyond $2.5$PN.
For a $2.5$PN-accurate expression of $\dot{\bar n}_0$ we only need to take into account the Newtonian order of the partial derivatives, since $\dot{\bar a}, \dot{\bar e} = \porder{5}$:
\begin{equation}
    \dot{\bar n} = \frac{\partial \bar n}{\partial \bar a} \dot{\bar a} + \frac{\partial \bar n}{\partial \bar e} \dot{\bar e} = - \frac{3 \dot{\bar a}}{2 \bar a^{5/2}} + \porder{6}.
\end{equation}
Hence, $\dot{\bar n}_0 =  \porder{5}$, and higher derivatives such as $\ddot{\bar n}_0 = \porder{10}$ can be safely neglected for a $2.5$PN local-in-time solution.
Integrating the mean motion then yields
\begin{equation}
\label{eq:sol_M}
    \bar M(t) = \bar M_0 + \bar n_0 t + \dot{\bar n}_0 \frac{t^2}{2} + \porder{6} \text.
\end{equation}
This expression is the strict PN-truncated secular solution for the mean anomaly, and is valid during time intervals over which the 2.5PN correction remains small.
The argument of the periapsis $\bar\omega$ is treated analogously, with $\dot {\bar\omega}_0 = \porder{2}$ and $\ddot {\bar\omega}_0 = \porder{7}$. This results in a power series truncated at linear order:
\begin{equation}
    \bar \omega(t) = \bar \omega_0 + \dot{\bar \omega}_0 t + \porder{6}, \qquad \dot{\bar \omega}_0 = \dot{\bar \omega}(\bar a_0, \bar e_0),
\end{equation}
where $\dot{\bar \omega} = \dot{\bar \omega}(\bar a_0, \bar e_0)$ is the conservative $1$--$2$PN expression given in Appendix~\ref{app:secular_conservative_dynamics}. Finally, $\bar\iota(t) = \bar \iota_0$ and $\bar \Omega(t) = \bar \Omega_0$.

\subsection{Complete dynamics}

We now reconstruct the complete non-averaged dynamics in the original variables.
Let $f$ be any phase-space quantity, expressed in terms of the variables used to solve the normal-form dynamics.
At $2.5$PN, the generator can be decomposed into conservative and dissipative parts according to Eq.\ \eqref{eq:gen_expansion}. Consequently, the Lie series splits into two parts
\begin{equation}
    \tilde f = [\mathcal T_{\mathfrak g} (\bar f_+)]_\PL
    = \mathcal T_g(\bar f)  + \varepsilon^5 \lPB \bar f^0_+, \mathfrak g^{2.5} \rPB_\PL + \porder{6}.
\end{equation}
The first term is the conservative part of the Lie series mapping (with conservative generator $g$) applied to the full dissipative normal-form solution. The second part introduces dissipative corrections to the normal-form mapping. 
Mixed contributions involving both the conservative generator and the $2.5$PN dissipative generator only enter beyond the accuracy considered here.

To isolate the conservative and dissipative contributions more explicitly, we split the normal-form solution itself as
\begin{equation}
    \bar f = \bar f^\mathrm{c} + \varepsilon^5 \bar f^{2.5} + \porder{6}.
\end{equation}
Here, $\bar f^\mathrm{c}$ collects the conservative pieces of the secular solution produced by the normal-form, while $\bar f^{2.5}$ denotes the corrections induced by the dissipative secular dynamics.
Inserting this decomposition into $\mathcal T_g(\bar f)$ while discarding higher-order corrections leads to:
\begin{equation}
    \tilde f = 
    \tilde f^\mathrm{c} + \varepsilon^5 \big( \bar f^{2.5} + \lPB \bar f^0_+, \mathfrak g^{2.5} \rPB_\PL \big) + \porder{6} \text.
\end{equation}
In this expression, $\tilde f^\mathrm{c} = \mathcal T_g(\bar f^\mathrm{c})$ denotes the conservative $2$PN Lie series evaluated on the conservative secular solution, or in other words, the complete conservative time evolution. The two remaining terms are the lead $2.5$PN dissipative corrections. The first one comes from the dissipative secular evolution in the normal form and therefore describes the slow inspiral. The second one comes from the dissipative part of the generator and restores the corresponding orbital-timescale corrections.
The conservative term $\tilde f^\mathrm{c}$ can eventually be replaced by a time evolution which does not originate from a Lie series computation, by appending the two leading dissipative corrections displayed above.

In particular, inserting the Delaunay variables allows us to recover their time evolutions:
\begin{subequations}
    \begin{align}
        \tilde M &= \tilde M^\mathrm{c} + \frac{1}{c^5} \pars*{ \bar M^{(2.5)} + \mathfrak g_L^{2.5} } ,\\
        \tilde \omega &= \,\tilde \omega^\mathrm{c} \,+ \frac{1}{c^5} \pars*{ \bar \omega^{(2.5)} + \mathfrak g_J^{2.5} } ,\\
        \tilde \Omega &= \,\tilde \Omega^\mathrm{c} \,+ \frac{1}{c^5} \pars*{ \bar \Omega^{(2.5)} + \mathfrak g_H^{2.5} } ,\\
        \tilde L &= \,\tilde L^\mathrm{c} \,+ \frac{1}{c^5} \pars*{ \bar L^{(2.5)} - \mathfrak g_M^{2.5} } ,\\
        \tilde J &= \,\tilde J^\mathrm{c} \,+ \frac{1}{c^5} \pars*{ \bar J^{(2.5)} - \mathfrak g_\omega^{2.5} } ,\\
        \tilde H &= \,\tilde H^\mathrm{c} \,+ \frac{1}{c^5} \pars*{ \bar H^{(2.5)} - \mathfrak g_\Omega^{2.5} } .
    \end{align}
\end{subequations}
The quantities $(\tilde M^\mathrm{c}, \tilde \omega^\mathrm{c}, \tilde \Omega^\mathrm{c}, \tilde L^\mathrm{c}, \tilde J^\mathrm{c}, \tilde H^\mathrm{c})$ denote the Delaunay variables evolved under the conservative $2$PN dynamics, which are known from the literature \cite[e.g.][]{Aykroyd2024}. The $2.5$PN secular corrections $(\bar M^{(2.5)}, \bar \omega^{(2.5)}, \bar \Omega^{(2.5)}, \bar L^{(2.5)}, \bar J^{(2.5)}, \bar H^{(2.5)})$ are computed in Sec.~\ref{sec:secular_dynamics}. In particular, Eqs.~\eqref{eq:a_e_sol}, \eqref{eq:sol_e}, and \eqref{eq:sol_M}, together with the relations $\bar L=\sqrt{\bar a}$, $\bar J=\sqrt{\bar a(1-\bar e^2)}$, and $H = J \cos\iota$, determine the dissipative secular corrections to $M$, $L$, $J$, and $H$. At this order the orbital plane does not undergo a secular precession, so that $\bar \omega^{(2.5)}=\bar \Omega^{(2.5)}=\bar \iota^{(2.5)}=0$. Finally, the dissipative generator contributions are obtained directly from Eq.~\eqref{eq:diss_gen}.


\section{Conclusion}

We have constructed a Lie perturbation scheme for nonconservative systems, building upon our double-phase-space Hamiltonian framework introduced in the previous work (\hampaper). We have shown that, by reframing the homological equation in the weak regime---sensitive only to the first order (in the virtual variables) structure of the Hamiltonian---we can mitigate the resonant obstructions inherent to a naive variable doubling and reduce the nonconservative problem to a set of decoupled, physical phase space homological equations [Eqs.\ \eqref{eq:hom_diss_ode}]. 
When combined with the Hamiltonian reconstruction methods of \hampaper, the Lie perturbation scheme can be straightforwardly adapted to a wide range of ordinary differential equations that are not initially given in symplectic (i.e., Hamiltonian) form. The method is applied in three key steps. First, the dissipative perturbation is encoded in the doubled Hamiltonian. Next, the Hamiltonian is linearly expanded with respect to virtual action-variables. Finally, the standard homological equation is solved independently for each coefficient via ordinary orbit averaging and primitive operations.
In the conservative limit [i.e., when $\mathcal R \to 0$ in Eq.~\eqref{eq:ham_diss_dec}], we have shown that the nonconservative homological equation reduces to the conservative case. This shows that our new framework agrees with the standard Lie-series construction, demonstrating a direct bridge between the conservative and dissipative settings.

As a demonstration, we have applied the method to solve the motion of radiating spinless binaries at $2.5$PN, in a fully canonical manner. The solution is expressed in osculating Delaunay elements using ADM coordinates. Accordingly, we construct the radiation‐reaction Hamiltonian and determine the Lie transform leading to a description of the secular dynamics. Our approach reproduces, simultaneously, the complete temporal evolution of the binary inspiral---including both the secular contributions (via the Peters–Mathews equations) and the oscillatory corrections.


These solutions could potentially be used as a first step to derive new accurate analytical templates in waveform applications for next-generation detectors such as the Laser Interferometer Space Antenna (LISA) \cite{LISA2017, LISA2024}, in regimes where a PN description is appropriate. This would require combining them with the appropriate multipolar wave-generation formalisms, dealing with hereditary contributions, and performing appropriate coordinate conversions.
The specific advantage of the Lie construction is that it keeps the secular inspiral and the oscillatory reconstruction in the same perturbative scheme; one may use the normal form dynamics alone when such averaged description is sufficient, or include the generator terms when corrections to waveform harmonics are needed. In long-duration observations such as for LISA, where small phase errors can accumulate over many cycles, this solution flexibility provides a controlled way, within the same analytical construction, to balance analytical precision against computational efficiency. A direct comparison with existing solutions at the same post-Newtonian order would then be required to determine whether the Lie method offers any practical accuracy advantages.

Our work also lays the groundwork for higher–order extensions. We expect the present construction to extend directly through the instantaneous radiation-reaction sector. Beyond the orders treated here, from $4$PN onward, hereditary tail contributions make the dynamics nonlocal in time, and complications arise.
Namely, in the present work, the conservative homological equation is solved by taking the average along the Newtonian flow to obtain the normal-form component, and then choosing the generator as an average-free primitive. In this setting, orbit averaging is exactly the natural projector of the Hamiltonian onto its resonant part. If the Hamiltonian is no longer a purely instantaneous function of phase space, as in the case of the $4$PN tails, a new approach for projecting the resonant dynamics must be found.

\section{Acknowledgements}
C.A. acknowledges the joint financial support of Centre National d'Études Spatiales (CNES) and École Doctorale Astronomie et Astrophysique d'Ile de France (ED127 AAIF). This work was also supported by the Programme National GRAM, by PNPS (CNRS/INSU), by INP and IN2P3 co-funded by CNES, and by CNES LISA grants at CEA/IRFU. 
We are very grateful to Shiladittya Debnath for their remarks and computation reviews.
We are thankful to A.\ Albouy, S.\ Bouquillon, and G.\ Faye for their constructive comments and discussions.

\bibliographystyle{apsrev4-2}
\bibliography{refs}

\newpage
\appendix

\section{ADM Hamiltonian perturbations up to 2PN}
\label{appendix:Hamiltonian_ADM} 

We recall the explicit expressions for the ADM Hamiltonian perturbations for a point-mass compact binary at 1PN and 2PN \cite{Schafer2024}:
\begin{subequations}
\begin{align}
    \mathcal H_1(\vec r, \vec \pi) &= \frac{1}{2 r^2} - \frac{1}{8} (1 - 3\nu) \pi^4  \nonumber\\&
    - \frac{1}{2 r} \Big[ (3 + \nu) \pi^2 + \nu (\uvec r \cdot \vec \pi)^2 \Big]  \text, \\
    \mathcal H_2(\vec r, \vec \pi) &= \frac{1}{16} \big( 1 - 5 \nu + 5 \nu^2 \big) \pi^6 + \frac{1}{8 r} 
    \Big[ \!\left( 5 - 20\nu - 3\nu^2 \right) \! \pi^4 \nonumber\\&
    - 2 \nu^2 \pi^2 (\uvec r \cdot \vec \pi)^2  - 3 \nu^2 (\uvec r \cdot \vec \pi)^4 \Big] \nonumber\\&
    + \frac{1}{2 r^2} \Big[ (5 + 8\nu) \pi^2 + 3\nu (\uvec r \cdot \vec \pi)^2 \Big] \nonumber\\&
    - \frac{1}{4 r^3} (1+3\nu) \text.
\end{align} 
\end{subequations}

\section{Secular conservative dynamics}
\label{app:secular_conservative_dynamics}

At conservative orders through $2$PN, the normal-form Hamiltonian is a function of the Delaunay actions only. For the nonspinning problem, rotational invariance removes any dependence on the nodal action $H$, so that
\begin{subequations}
\begin{align}
    \frac{\dd \bar M}{\dd t} &= \frac{\partial \mathcal H^{\mathrlap{*,\mathrm{cons}}}}{\partial L} \phantom{\null^{*,\mathrm{cons}}} \equiv \bar n \text, &    
    \frac{\dd \bar L}{\dd t} &= 0 \text, \\
    \frac{\dd \bar \omega}{\dd t} &= \frac{\partial \mathcal H^{\mathrlap{*,\mathrm{cons}}}}{\partial J} \phantom{\null^{*,\mathrm{cons}}} \equiv \dot{\bar \omega} \text, &    
    \frac{\dd \bar J}{\dd t} &= 0 \text, \\
    \frac{\dd \bar \Omega}{\dd t} &= \frac{\partial \mathcal H^{\mathrlap{*,\mathrm{cons}}}}{\partial H} \phantom{\null^{*,\mathrm{cons}}} =0 \text, &
    \frac{\dd \bar H}{\dd t} &= 0 \text,
\end{align}
\end{subequations}
The conservative secular dynamics therefore consists of the mean motion and the periastron advance frequency. Explicitly, in the same rescaled ADM variables used throughout the paper,
\begin{subequations}\label{eq:conservative_secular_frequencies}
\begin{align}
     \bar n &= \frac{1}{\bar L^3} 
     + \frac{1}{c^2} \Bigg\{ \frac{9}{\bar J \bar L^4}+\frac{\nu -15}{2 \bar L^5} \Bigg\} 
     + \frac{1}{c^4} \Bigg\{ \frac{15 (7 - 2 \nu)}{4 \bar J^3 \bar L^4} \nonumber\\& \quad 
     +\frac{54}{\bar J^2 \bar L^5}+\frac{60 \nu - 525}{4 \bar J \bar L^6}+\frac{3 \left(\nu ^2-15 \nu +145\right)}{8 \bar L^7} \Bigg\} \text, \\
    \dot{\bar \omega} &= \frac{1}{c^2} \frac{3}{\bar J^2 \bar L^3} 
    + \frac{1}{c^4} \Bigg\{ \frac{15 (7 - 2 \nu)}{4 \bar J^4 \bar L^3}+\frac{27}{\bar J^3 \bar L^4} \nonumber \\& \quad
    +\frac{12 \nu - 105}{4 \bar J^2 \bar L^5}\Bigg\} \text.
\end{align}
\end{subequations}
Alternatively, these frequencies can be expressed in terms of the semimajor axis and eccentricity using the relations
\begin{equation}
    \bar L=\sqrt{\bar a}\text, \qquad
    \bar J=\sqrt{\bar a(1-\bar e^2)}\text.
\end{equation}

\section{Kepler flow integrals} \label{appendix:flow_integrals}

\subsection{Primitive functions}

Building on the results of \cite{Aykroyd2024}, we construct additional explicit primitive functions (antiderivatives) along the unperturbed flow of the Keplerian Hamiltonian $\mathcal H^{0}$. Throughout, $\Phi_{t}^{\mathcal H^{0}}$ denotes this flow. For any phase-space function $f$ we introduce the average-free primitive
\begin{equation}
    \primitive[f] = \int f \circ \Phi_t^{\mathcal H^0} \, \dd t, \quad \avg{\primitive[f]} = 0\text,
\end{equation}
where $\avg{\argdot}$ stands for the orbital average over one Keplerian period:
\begin{equation}
    \avg{f} = \frac{1}{T} \int_{-T/2}^{T/2} f  \circ \Phi_t^{\mathcal H^0} \, \dd t \text.
\end{equation}

\paragraph*{Radial primitives}
Because $\{r, \mathcal H_0 \} = p_r$, every function $f$ depending solely on the magnitude of the separation admits the immediate identity,
\begin{equation}
    \primitive[p_r f'(r)] = f(r) - \avg{f(r)} \text. \label{eq:radial_primitive}
\end{equation}

\paragraph*{Angular primitives}
Resolution of the dissipative homological equations for the $2.5$PN dynamics further requires primitives of the three classical orbital anomalies.
In the following anomaly primitives, the orbital angles are understood on a fixed continuous branch over one radial period, which we take as $M \in (-\pi, \pi]$, with $u(M)$ and $v(M)$ chosen continuously and satisfying $u(0) = v(0) = 0$.

At the Newtonian level, the mean anomaly evolves linearly as $M = M_0 + t/L^3$, yielding for all $k \in \naturals$,
\begin{equation}
    \primitive[M^k] = \frac{L^3}{(k+1)} \left( M^{k+1} - \avg{M^{k+1}} \right) \text.
\end{equation}
Accordingly, the osculating eccentric anomaly can be defined in phase-space via Kepler's law \cite{Aykroyd2024} as $u = M + r p_r / L$, so that Eq.\eqref{eq:radial_primitive} gives:
\begin{equation}
    \primitive[u] = \frac{M^2 L^3}{2}+\frac{r^2}{2 L} - K \text,
\end{equation}
where the integration constant cancels out the average:
\begin{equation}
    K = \frac{L}{4} \Big(5L^2 - 3J^2 \Big) + \frac{\pi^2 L^3}{6} \text.
\end{equation}

The nontrivial case---our main result in this appendix---is that of the true anomaly $v$. After 
transforming the integration variable from $t$ to $v$ we obtain:
\begin{equation}
    \primitive[v] = J^3 \int \frac{v}{(e \cos (v)+1)^2} \dd v \, \text,
\end{equation}
whose closed form involves non-elementary functions. We begin by performing a standard half-angle substitution on the orbital anomalies:
\begin{align}
v &= 2 \arctan(x) \text, & u &= 2 \arctan \left(\frac{x}{\beta }\right) \text,
\end{align}
with $\beta = \sqrt{{(1+e)}/{(1-e)}}$.
Substitution reduces the primitive to the form
\begin{equation}
    \primitive[v] = \frac{4 J^3}{(1-e)^2} \int \frac{\left(x^2+1\right)}{\left(\beta ^2+x^2\right)^2} \arctan(x) \,\dd x \text.
\end{equation}
The rational polynomial kernel can be split via partial fractions yielding the two pieces 
\begin{equation}
    \primitive[v] = \frac{4 J^3}{(1-e)^2}
    \big( \primitive_1[v] + \primitive_2[v] \big) - \avg{\primitive[v]} \text,
\end{equation}
where
\begin{align}
    \primitive_1[v] &= \int \frac{\beta^2 - 1}{2 \beta^2} \frac{x^2-\beta^2}{\pars*{x^2+\beta^2}^2} \arctan(x) \,\dd x \text, \\
    \primitive_2[v] &= \int \frac{\beta^2 + 1}{2 \beta^2} \frac{1}{x^2+\beta^2} \arctan(x) \,\dd x \text.
\end{align}
The first piece can be evaluated via standard methods, and equals:
\begin{equation}
    \primitive_1[v] = \frac{1}{4 \beta^2} \pars*{\ln{\pars*{\frac{x^2 + 1}{x^2 + \beta^2}}} - \frac{2 x \pars*{\beta^2-1}}{x^2 + \beta^2} \arctan{(x)} } \text.
\end{equation}
The second integral requires the dilogarithm function, defined as the analytic continuation of the complex series
\begin{equation}
    \dilog(z) = \sum_{k=1}^\infty \frac{z^{k}}{k^2}, \quad |z| < 1, \label{eq:Li2_series}
\end{equation}
namely
\begin{equation}
    \dilog(z) = - \int_0^z \frac{1}{u} \ln{\pars{1-u}} \,\dd u, \quad z \in \complexes \,\textbackslash\, [1, \infty) \text. \label{eq:dilog_int}
\end{equation}
The dilogarithm has a branch cut in the real interval $[1, \infty)$, and satisfies the complex conjugate relationship
\begin{equation}
    \dilog(z^\dagger) = \dilog^\dagger(z) \text.
\end{equation}
We introduce a special function $\xi : \reals \to \reals$ representing the real part of $\dilog$ in polar complex coordinates, which satisfies
\begin{equation}
    \xi(\theta) = \frac{1}{2} \pars*{ \dilog(\rho \ee^{\ii \theta}) + \dilog(\rho \ee^{- \ii \theta}) } = \Re { \left\{ \dilog(\rho \ee^{\ii \theta}) \right\} } \text.
\end{equation}
For $\rho \in [0, 1)$ the arguments of $\dilog$ do not cross the branch cut, so that $\xi$ remains everywhere analytic.

The derivative of $\xi$ with respect to the polar angle $\theta$ can be evaluated with aid of the integral expression [Eq.\ \eqref{eq:dilog_int}], so that
\begin{equation}
    \xi'(\theta) = \Im{\big\{ \ln{(1 - \rho \ee^{\ii \theta})} \big\}} = - \arctan{ \pars*{\frac{\rho \sin\theta}{1 - \rho \cos\theta}} }\text.
\end{equation}
Since $\xi$ is continuous and $2\pi$-periodic, $\xi'$ must average to zero; hence the inverse tangent must be evaluated in the principal branch for all $\theta$.
Introducing the half-angle substitution $y = \tan{(\theta/2)}$
and applying the elementary trigonometric identity
\begin{equation}
    \arctan{\pars*{\frac{a - b}{1+ a b}}} \equiv \arctan{a} - \arctan{b}  \text,
\end{equation}
which is valid whenever both sides lie in the same branch, we recover the equivalent expression
\begin{equation}
    \xi'(\theta) = \frac{\theta}{2} - \arctan{\pars*{\frac{1+\rho}{1-\rho} \tan{\pars*{\frac{\theta}{2}}}}} \text.
\end{equation}
We can now identify $\theta = u$ and $\rho = (\beta-1)/(\beta+1) < 1$, so that $\xi'(u) = (u - v)/2$. Applying the chain rule, we compute the derivative of $\xi$ with respect to the variable $x$,
\begin{equation}
    \frac{\dd \xi}{\dd x} = \frac{2 \beta}{x^2 + \beta^2} \left( \arctan{\pars*{\frac{x}{\beta}}} - \arctan{\pars*{x}} \right)  .
\end{equation}
Integrating again in $x$ allows us to identify the second primitive,
\begin{equation}
    \primitive_2[v] = \frac{\beta ^2+1}{4 \beta ^3} \pars*{ \arctan^2{ \pars*{\frac{x}{\beta }}} - \xi(u) } \text.
\end{equation}
Next, averaging the series representation in Eq.~\eqref{eq:Li2_series} term-wise gives
\begin{equation}
    \!\!\avg{\xi} = \sum_{k=1}^\infty \frac{\rho^k}{k^2} \avg{ \cos{(k u)} } = - \frac{e \rho}{2} = \frac{1}{2} \left( \frac{J}{L} - 1 \right) \text,
\end{equation}
since only the $k=1$ harmonic survives.

Collecting all pieces, enforcing the average-free condition, and writing everything in terms of orbital parameters we arrive at the main result:
\begin{multline}
    \primitive[v - u] = 
    J L^2 \ln {\pars*{ \frac{2r}{ L (J+L)} }} 
    - 2 L^3 (\xi - \avg{\xi}) 
    -L r \\
    -L^2 (v - u) r p_r     
    + \frac{1}{4} \big( 9 L^3 - 6 J L^2 + J^2 L \big)  \text.
\end{multline}

For completeness we also compute the time derivative of $\xi(u)$,
\begin{equation}
    \frac{\dd \xi}{\dd t} = \frac{1}{2 L} \frac{u - v}{r} \text,
\end{equation}
so that we obtain for free
\begin{equation}
    \primitive\left[ \frac{u - v}{r} \right] = 2 L \big( \xi - \avg{\xi} \big) \text.
\end{equation}

\begin{widetext}
\subsection{Logarithmic average}
In this section we evaluate the time‐averaged logarithm of the radial distance in a Keplerian flow. Namely, the aim is to compute
\begin{equation}
    \avg{\ln r} = \frac{1}{T} \int_{-T/2}^{T/2} \ln r \circ \Phi^{\mathcal H^0}_t \, \dd t \text.
\end{equation}
We begin à la Feynman by introducing a parameter into the integral via a function $H$ where
\begin{equation}
    H(x) = \frac{1}{2 \pi} \int_{-\pi}^{\pi} \left(1 - e \cos u \right) \ln \Big( L^2 \left(1 - e x \cos u \right) \Big) \, \dd u \text,
\end{equation}
so that $H(1) = \avg{\ln r}$ is our target integral. As a preliminary step, we note that the evaluation at $x=0$ is straightforward:
\begin{equation}
    H(0) = 2\ln L \text.
\end{equation}
To compute $H(x)$ in general, differentiation under the integral sign with respect to $x$ yields
\begin{equation}
    H'(x) = - \frac{1}{2\pi}\int_{-\pi}^{\pi} \frac{\left(1 - e \cos u \right)}{\left(1 - x e \cos u \right)} e \cos u \, \dd u  = \frac{(1-x)(1-\sqrt{1- e^2 x^2})}{x^2 \sqrt{1 - e^2 x^2} } \text.
\end{equation}
Integrating the above expression with respect to $x$, we obtain
\begin{equation}
    H(x) = \left( \frac{1}{x} - \frac{\sqrt{1-e^2 x^2}}{x} + 2 \ln x - \ln\big(1 - \sqrt{1 - e^2 x^2} \big) \right) + k \text.
\end{equation}
where $k$ is an integration constant. The constant is determined by taking the limit as $x \to 0$, utilising the continuity of the integral over the interval $(0,1]$. Matching the limit to the known value $H(0)=2\ln L$, we find $k = \ln (e^2 L^2 / 2)$. Hence
\begin{equation}
    \avg{\ln r} = H(1) 
    = 1 - \frac{J}{L} + \ln \left( \frac{L (J + L)}{2} \right) \text.
\end{equation}

\end{widetext}

\end{document}